\documentstyle[12pt,epsfig]{article}
\begin{document}

\newcommand{\fig}[2]{\epsfxsize=#1\epsfbox{#2}}

\begin{titlepage}
\vspace*{2.cm}
\begin{center}
{\Large \bf Exact renormalization group}
\\[.2cm] {\Large \bf and applications to disordered problems: part I}
\\[1cm] \today
\\[1cm]{\large Pascal Chauve}\footnote{E-mail: chauve@lpt.ens.fr}
\\{\em \small Laboratoire de Physique des Solides}
\\ {\em \small Universit{\'e} Paris-Sud, B{\^a}t. 510}
\\ {\em \small 91405 Orsay - France}
\\[1cm]{\large Pierre Le Doussal}\footnote{E-mail: ledou@lpt.ens.fr}
\\{\em \small Laboratoire de Physique Th{\'e}orique}
\\ {\em \small {\'E}cole Normale Sup{\'e}rieure}
\\ {\em \small 24, rue Lhomond}
\\ {\em \small 75231 Paris C{\'e}dex 05 - France}
\\[1.cm]
\end{center}
\begin{abstract}
We develop a systematic multi-local expansion
of the Polchinski-Wilson exact renormalization group (ERG) equation.
Integrating out explicitly the non local interactions, we reduce the 
ERG equation obeyed by the full interaction functional 
to a flow equation for a function, its local part.
This is done perturbatively around fixed points, but exactly to
any given order in the local part. It is thus controlled, at variance with
projection methods, e.g. derivative expansions or local potential
approximations. 
Our method is well-suited to problems such as the
pinning of disordered elastic systems, previously described via functional
renormalization group (FRG) approach based on a hard cutoff scheme.
Since it involves arbitrary cutoff functions, we explicitly 
verify universality to ${\cal O} (\epsilon =4-D)$, 
both of the $T=0$ FRG equation and
of correlations. Extension to finite temperature $T$ yields the
finite size ($L$) susceptibility fluctuations characterizing mesoscopic 
behaviour $\overline{ (\Delta \chi )^2 } \sim L^\theta/T$, where
$\theta$ is the energy 
exponent. Finally, we obtain the universal scaling function 
to ${\cal O}(\epsilon^{1/3})$ 
which describes the ground state of a domain wall
in a random field confined by a field gradient, compare with
exact results and variational method. Explicit two loop exact RG
equations are derived and the application to the FRG problem
is sketched. 
\end{abstract}
\end{titlepage}

\section{Introduction}

The idea of writing an exact equation for the scale dependence
of the full action functional already appears in the review
of Wilson and Kogut \cite{wilson_kogut}. Since it is an equation for a full
functional of the fields, its detailed analysis is hindered by 
technical complications. The much simpler Wilson momentum shell
\cite{wilson_kogut}
integration method is commonly used for one loop calculations.
Since it does not follow the full functional, subsequent efforts
were made to embed it into a better controlled sharp cutoff exact RG 
\cite{wegner_houghton,morijmp}. For practical perturbative calculations
beyond one loop, field theoretical renormalization methods are more
often used since they have proved vastly more efficient. 
However, the exact RG equations offer the hope to develop ab initio
calculation relying on no assumption, possibly non perturbative, from 
any bare model.
In principle it should be useful to obtain precise results when
applied to bare theories for which we have little insight on 
possible underlying field theoretical description.

In the work of Polchinski \cite{polchinski}, the exact RG
equation was put on a more precise and aesthetic framework,
and used to prove the renormalizability of the $\phi ^{4}$ theory 
in four dimensions. The exact renormalization group equations
indeed provide formal results or general proofs about symmetries 
\cite{morris_symmetry}. For practical calculations however, one needs
to truncate in some way these highly complicated functional
equations. To do so, different  procedures have been
proposed \cite{Wet91-1,hasenfratz^2}, and have been mainly
applied to the study of non-perturbative problems 
\cite{revue}. For example the exponents of the $O(n)$ model in three
dimensions were estimated \cite{GRN} using a choice of
truncation. One commonly used projection method is the so called local
potential approximation \cite{NCS1}, obtained by 
a constant background field method 
neglecting the momentum dependence.
Further extensions include additional projections on higher gradients
of the field \cite{morrisderiv}. Although very interesting, these
projection methods are often uncontrolled. More accurate results
are expected if more couplings are kept, which is possible with
heavy numerical integrations of flow equations. In this respect, exact RG
as a tool is now used both in particle and condensed matter physics.
For instance, outstanding problems in strongly correlated electrons 
such as the Hubbard model in $D=2$, have been recently studied by numerically
integrating the flow of a large number of vertices\cite{hubbard},
using a fermionic version of the Polchinski
equation\cite{polchinski}. 

By contrast, comparatively a few works use exact RG
method to develop {\it perturbative} calculations. One example is the  
computation of the beta function of $\phi^4$ in $4-\epsilon$ dimensions to one
loop\cite{hughes_liu}, where universality is made particularly
explicit through the use of an arbitrary cutoff
function. Although obviously more
powerful methods are available in that case, there are some problems 
in condensed matter physics which appear within reach of perturbative
calculations but for which no coherent field theoretical formulation
is available at present. This is the case for the pinning of an
elastic system in a random potential, for 
which a momentum shell RG method has been
developed\cite{fisher_functional_rg}. 
In this problem, an infinite number of coupling constants becomes relevant for
$D < D_c=4$ and one must write a RG equation for a full function $R(u)$ (the
second cumulant of the disorder), hence the name ``functional'' 
renormalization group (FRG). As such it differs from standard field
theoretical RG. Thus, to understand better this problem, i.e to show
explicitly universality to one loop and beyond, there is a need for 
a perturbatively controlled exact RG method, able to admit a full
function, the local part $R(u)$ as a small parameter. Indeed the
field theoretical formulation is frought with difficulties, in particular
because the function $R(u)$ develops non analytic behaviour
at finite scale. These issues are discussed 
in a related work \cite{kay_dynamics}. 

In this paper we develop a novel method to solve the Polchinski exact
renormalization  group equation and use it for explicit calculations.
Writing the action as a sum of multilocal interactions,
we note that the Polchinski equation naturally reduces to a hierarchy
of equations obeyed by simple functions. This hierarchy can be solved
in an expansion  in powers of the local part. Indeed, we find that
exact integration of the multilocal parts yields a single RG equation
for the local part. The method is thus controlled around fixed points
where the local part is proportional to a well defined small parameter
(e.g. $\epsilon =4-D$). It does not require any arbitrary projection
procedure or neglect of operators, as is usually done in derivative
expansions or local potential approximations. In addition, we obtain
explicit formulas for any correlation function which allow 
for practical calculations. Since this is
done for arbitrary cutoff functions, it allows explicit check of universality
order by order in the expansion. 

The aim of this paper is twofold. On the one hand we present the general method to all
orders, valid for a large class of theories. We derive the explicit form of
the exact RG equation for the local interaction up to third order. 
On the other hand we apply this method to several problems, first as a check,
to the $O(n)$ model, and second to the FRG for disordered elastic systems.
Explicit calculations and applications in this paper are restricted mainly to
one loop. Although briefly mentionned here, applications to two loops will
be detailed in a companion paper \cite{us_part2}.

Two variants of the method are presented. The most direct one consists in
a straight expansion of the action in multilocal terms.
The second one consists instead of first absorbing tadpoles into the
interaction (so-called Wick ordering), then expanding. Being
inequivalent, they provide independent checks of the universal results.
The first method yields more complicated equations but, 
can be better suited to some problems, such as the $T>0$ FRG. 
Note that although 
Wick ordered versions of the Polchinski equation have been studied
before, the multilocal expansion performed here is to our knowledge
novel.

As mentionned above, the method is indeed well-suited to the FRG 
for disordered elastic systems of internal dimension $D$ since there
the full local part is controlled by $\epsilon =4-D$.
It allows us to show that the one loop FRG
equation, as well as correlation functions, are 
independent of the cutoff function. In addition we obtain 
higher cumulants of the renormalized disorder, which 
as the second cumulant, are non analytic functions.
This is necessary to escape the so-called dimensional
reduction \cite{dimred2}, i.e the property of the present theory by
which all perturbative calculations at $T=0$ are identical to
the same calculation in a trivial gaussian theory
\cite{dimred1} (see Appendix \ref{fleurs}).
This nonanalytic behaviour is rounded at finite temperature $T$
and we obtain the scaling form of the rounding region.
This allows us to compute, for the first time using the FRG method,
the susceptibility fluctuations which characterize the glassy
behaviour of finite size systems.
Finally, we obtain the universal 
${\cal O}(\epsilon^{1/3})$ correlation function 
which describes the ground state of a domain wall
in a random field confined by a field gradient, compare with
exact results and variational method.

The method presented in this paper also allows to investigate the
theory of disordered elastic systems beyond lowest order in $\epsilon$ (one loop).
A recent two loop calculation was presented in \cite{larkin_2loops}.
However since it was performed at $T=0$, and for analytic $R(u)$
it fails beyond a finite length (Larkin length) and cannot describe
universal properties. The application of exact FRG to next order
is described in \cite{us_part2}. We sketch here however 
some preliminary results.

The paper is organized as follows. In Section~\ref{exactrgpro}
we present in a pedagogical way the conventional exact RG method.
Appendix~\ref{invariance} and Appendix~\ref{examples} provide
complements, respectively about general invariance properties
of the correlations and about examples of solvable cases of
the Polchinski equation. In \ref{multiloc}, the multilocal expansion in the
local part $U$
is introduced up to bilocal terms. The ensuing RG equations to order $U^2$
are given in~\ref{sollow}. The multilocal expansion to arbitrary order
and the RG equation to order $U^3$ is given in Appendix~\ref{higher}.
The multilocal expansion of the Wick ordered functional
up to second order, and the resulting one loop RG equation is
presented in \ref{modpol}. The general multilocal expansion 
(Appendix~\ref{vhatmulti}) and the resulting form to third order is given in
Appendix~\ref{vcube}. The explicit two loop RG equation is obtained
in \ref{loopexpansion}. Application to the $O(n)$ model to one loop
is presented in~\ref{on}. We then turn to applications to disordered
elastic systems in Section~\ref{elastic} (one loop) and 
Section~\ref{2l} (two loops).
First we recall and generalize in
Appendix~\ref{fleurs} the dimensional 
reduction phenomenon. Then the $T=0$ FRG equations are
established in Section~\ref{frgequniv} and finite temperature
extension are given 
in Section~\ref{finitetemp}. Finally, the calculations of the 
scaling function in the random field Ising model is performed in~\ref{biased}.
We sketch in Appendix~\ref{variational} the
variational calculation to be compared with the FRG results of
\ref{biased} and sketch some preliminary steps of
a two loop FRG in Appendix~\ref{totwoloops}.

\section{Method}
\label{method}

\subsection{Exact RG procedure}
\label{exactrgpro}

Consider a system whose state is 
described by a bosonic {\it field} $\phi^{x} _{i}=\phi _{i}(x)$, where $x$
denotes position in space, and $i$ is a
general label denoting e.g. fields indices, spin, replica indices,
additional coordinate (e.g. time) etc. (or more generally any
quantity which will not undergo the coarse-graining). 
The system, in the presence of external {\it sources} $J_{i}^{x}$, 
is described by the partition function:
\begin{eqnarray}
Z (J)=\int_{\phi } e^{J:\phi -{\cal S} (\phi)}
\end{eqnarray}
obtained by the integration over the field $\phi $,
where the {\it action} ${\cal S} (\phi )$ is a functional of the field $\phi
$, and $J:\phi $ denotes here and in the following the full scalar product
(e.g. $\int_{x} \sum_{i} J_{i}^{x} \phi_{i}^{x}$, with $\int_{x}
\equiv \int d^{D}x$). In a problem of
equilibrium statistical mechanics, ${\cal S} (\phi )={\cal H} (\phi
)/T$ where ${\cal H} (\phi )$ is the hamiltonian and $T$ the
thermodynamic temperature, the free energy being $F=-T\ln Z (0)$.
Averages of any  
{\it observable} ${\cal A} (\phi )$ (i.e. functional of $\phi $) are
defined by
\begin{eqnarray}
\langle {\cal A} (\phi)\rangle_{{\cal  S}} =\frac{\int_{\phi } {\cal A} (\phi
) e^{-{\cal S} (\phi )}}{\int_{\phi } e^{-{\cal S} (\phi )}}
\end{eqnarray}

\begin{figure}[hbt]
\centerline{\epsfig{file=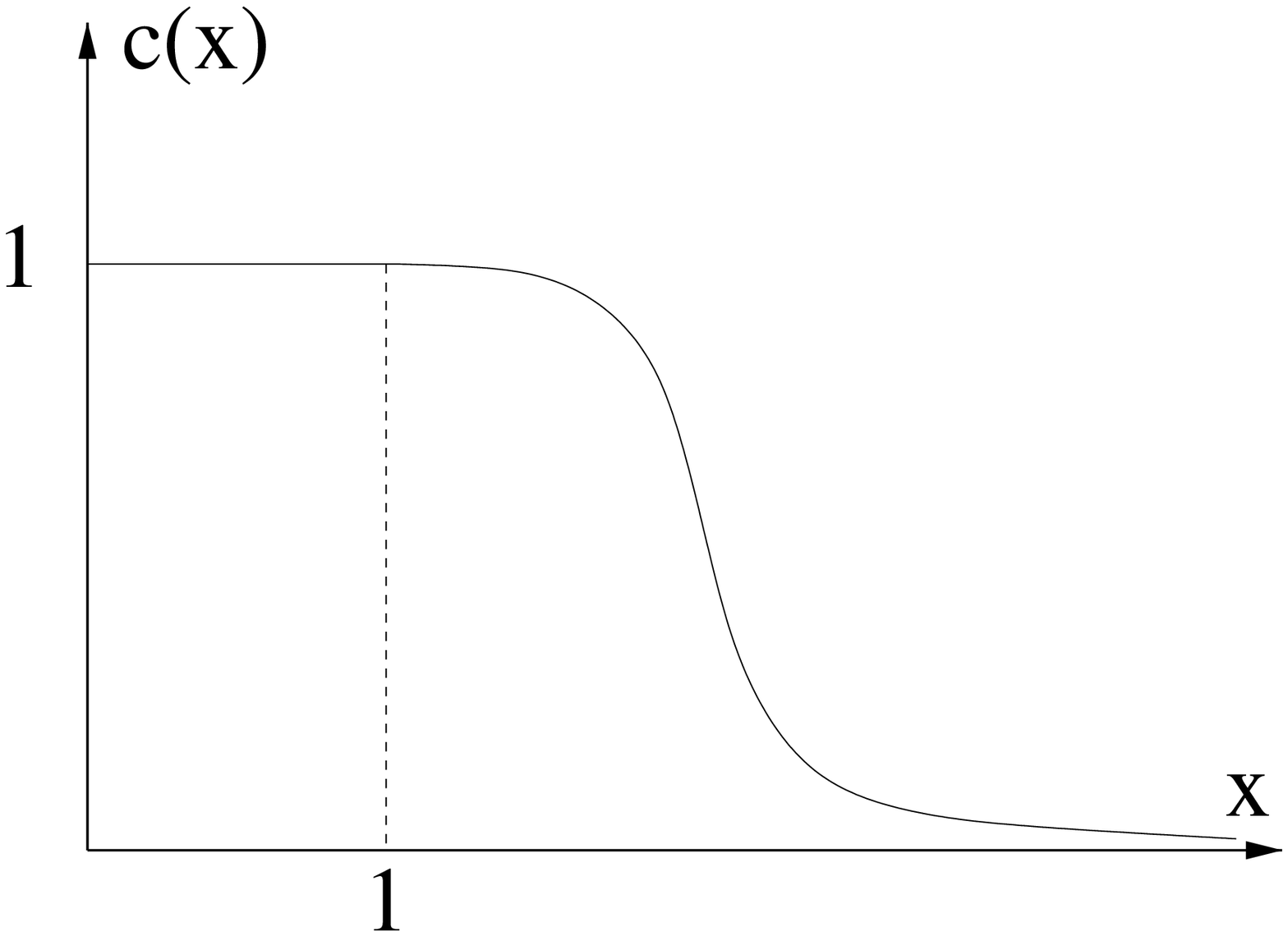,angle=0,width=5.cm} 
\epsfig{file=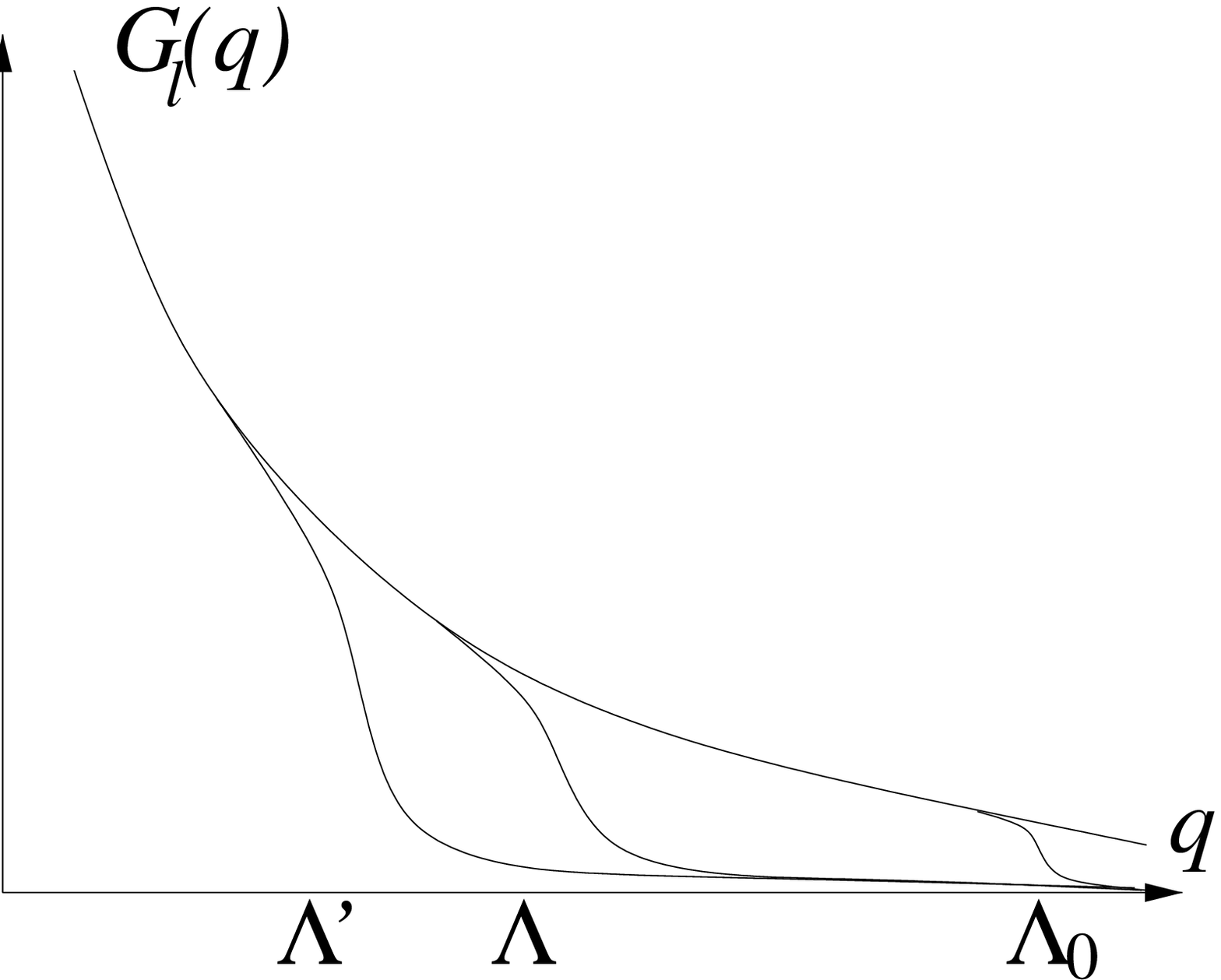,angle=0,width=5.cm}}
\caption{Cutoff function.}
\label{coupure}
\end{figure}

The usual way to compute correlation functions and averages is to perform a
perturbation expansion, writing the action as a sum of a quadratic
part and a non-linear part ${\cal V} (\phi)$ 
\begin{equation}\label{generalaction}
{\cal S} (\phi )=\frac{1}{2}\phi :G^{-1}:\phi + {\cal  V} (\phi)
\end{equation}
where ${\cal  V} (\phi)$ a functional of $\phi $ and $G_{ij}^{xy}
=G_{ji}^{yx}$ is a symmetric invertible matrix,
$\phi :G^{-1}:\phi=\sum_{ij} \int_{xy} 
\phi_{i}^{x} (G^{-1})_{ij}^{xy} \phi_{j}^{y} $. In the following we
denote the Gaussian average of any observable by
$[{\cal A} (\phi )]_{G}=\langle {\cal A}
(\phi )  \rangle _{{\cal S}_{G}}$ with respect to the quadratic theory 
${\cal S}_{G}=\frac{1}{2}\phi :G^{-1}:\phi $.
We introduce the generating function of all
correlation functions
\begin{equation}
W (J)=\ln \left[ e^{J:\phi -{\cal V} (\phi )}\right]_{G}
\end{equation}
Note that it differs from the usual definition by a $J$-independent
quantity $\ln Z (J) =W (J)+\frac{1}{2}{\rm Tr}\ln G$. 
The ultra-violet cutoff, present in physical models, is necessary to
yield finite results in the perturbative calculation with respect to
${\cal V}$. A broad class of soft cutoffs can be 
implemented on the Gaussian part, giving a vanishing weight
to fast fields. For example a
scalar massive theory, rotationnally invariant, 
is regularized in the UV by the following
general cutoff function
\begin{equation}\label{examplecutoff}
G (q)=\frac{c (\frac{q^{2}}{2\Lambda ^{2}})}{q^{2}+m^{2}}
\end{equation}
where $c (0)=1$ and $c (s)$ decreases rapidly to zero for $s>1$ as in
Figure~\ref{coupure}.

The exact RG method \cite{wilson_kogut,polchinski}
consists in varying the
cutoff $\Lambda $ and writing an equation for the function ${\cal V} (\phi )$ 
so as to conserve exactly the averages of all observables
involving only ``slow'' modes of the field. More precisely, the
average of an observable ${\cal A} (\phi )$ depending only on modes
$q<\Lambda '$ of the field $\phi $ can be computed within any of the
theories linked by the equation presented by Polchinski 
corresponding to a cutoff
$\Lambda >\Lambda '$. As in any RG procedure, the strategy will be to
compute averages of slow observables using the coarse-grained theory
of cutoff $\Lambda '\ll \Lambda _{0}$. 

To this aim, a set of actions ${\cal S}_{l}$ 
\begin{equation}
{\cal S}_{l}(\phi)=\frac{1}{2}\phi :G_{l}^{-1}:\phi +{\cal V}_{l} (\phi)
\end{equation}
is introduced, where the Gaussian part is 
an arbitrary function $G_{l}$ of $l$ (e.g. corresponding to a
cutoff $\Lambda _{l}=\Lambda _{0}e^{-l}$). The initial
propagator, corresponding to a cutoff $\Lambda _{0}$, 
is denoted $G_{l=0}\equiv G$ 
and the bare interaction ${\cal V}_{l=0}
\equiv {\cal V}$. The correlation functions in ${\cal S}_{l}$ derive
from $W_{l} (J)=\ln \left[e ^{J:\phi -{\cal V}_{l} (\phi )}
\right]_{G_{l}}=\ln Z_{l} (J)-\frac{1}{2}{\rm Tr}\ln G_{l}$. 

For any given $\Lambda '$, one defines a ($\Lambda '$--)slow observable to be
a functional of $\phi $ depending only on the $\phi ^{q}=\phi (q)$ with
$q<\Lambda '=\Lambda_{0} e^{-l'}$.
We want to choose the $l$-dependent 
non-linear part ${\cal V}_{l} (\phi )$ so that the averages of 
slow observables remain unchanged.
Through differentiation, it is equivalent to ensure that
\begin{equation}\label{ejphi}
\partial_{l} W_{l} (J) \mbox{ independent of }J
\end{equation}
for {\it any} source $J$ with $J^{q}=0$ for $q>\Lambda '$. Using the
general identity 
\begin{equation}\label{gauss}
\partial_{l}[{\cal A}_{l} (\phi )
]_{G_{l}}=\frac{1}{2}{\rm Tr}\left(\partial_{l}G_{l}:[
\frac{\delta ^{2}}{\delta \phi \delta \phi }{\cal A}_{l} (\phi )
]_{G_{l}} \right)+[ \partial_{l} {\cal A}_{l} (\phi )
]_{G_{l}}
\end{equation}
valid for Gaussian averages of any $l$-dependent observable 
${\cal A}_{l} (\phi )$ and applying it to 
${\cal A}_{l} (\phi )=e^{J:\phi -{\cal V}_{l} (\phi )}$, one finds
\begin{eqnarray}
&&\partial_{l} [e^{ J:\phi -{\cal V}_{l} (\phi
)}]_{G_{l}}=\nonumber\\
&&\left[\left(\frac{1}{2}(J-\frac{\delta }{\delta \phi
}{\cal V}_{l} (\phi )): \partial_{l}G_{l}:(J-\frac{\delta }{\delta \phi
}{\cal V}_{l} (\phi ))-\partial_{l}{\cal V}_{l} (\phi )-\frac{1}{2}{\rm
Tr}\partial_{l}G_{l}:\frac{\delta ^{2}}{\delta \phi \delta \phi}{\cal
V} (\phi ) \right)e^{ J:\phi -{\cal
V}_{l} (\phi )} \right]_{G_{l}} \nonumber 
\end{eqnarray}
where here and in (\ref{gauss}), ${\rm
Tr} A:B\equiv \sum_{ij}\int_{xy}A_{ij}^{xy} B_{ji}^{yx} $. 
Hence, if ${\cal V}_{l} (\phi )$ satisfies the Polchinski functional equation
\begin{equation}\label{poleq}
\partial_{l}{\cal V}_{l} (\phi )=-\frac{1}{2}{\rm Tr}\left(
\partial_{l}G_{l}:\frac{\delta ^{2}}{\delta \phi \delta \phi }{\cal V}_{l}
(\phi )\right) +  \frac{1}{2} \frac{\delta}{\delta \phi}{\cal V}_{l}(\phi
):\partial_{l}G_{l}:\frac{\delta}{\delta \phi}{\cal V}_{l}(\phi )
\end{equation}
then the above conservation condition (\ref{ejphi}) is satisfied. We
have used explicitely the condition
\begin{equation}
J:\partial_{l}G_{l}=0
\end{equation}
which imposes that the cutoff function verifies
$\partial_{l}G_{l}^{q} =0$ for $q<\Lambda '$ and 
$l>l'$. Hence, for the example (\ref{examplecutoff}) 
one has to choose~\cite{analyticity} cutoff
functions $c (s)$ such that $c (s)= 1$ for $0\leq s\leq s_{0}$
with some (arbitrary) $s_{0}$.

The above framework is in fact too restrictive. We can easily 
lift the restriction on slow modes (and on the form of the cutoff 
function $c(s)$). The applications of Polchinski equation can be generalized
to the computation of {\it any} observable (not restricted to 
be ``slow''). As shown in Appendix~\ref{coarsegrain}
one can indeed express $W (J)$ in terms of any of the
$l$-dependent actions ${\cal S}_{l} (\phi) $:
\begin{eqnarray}\label{wwj}
W (J)=\frac{1}{2}J:(G-G:G_{l}^{-1}:G):J+W_{l} (J:G:G_{l}^{-1})
\end{eqnarray}
In fact we show in Appendix~\ref{zphi} an even more 
general method which allows for arbitrary field rescalings.

Differentiating $W(J)$ once yields $\langle \phi \rangle _{\cal
S}=G:G_{l}^{-1}\langle \phi \rangle _{{\cal S}_{l}}$, and once again 
yields the two point connected correlation
function
\begin{equation}\label{phiphi}
\langle \phi \phi \rangle^{c} _{{\cal S}} = G +
G:G_{l}^{-1}:\left( \langle \phi \phi \rangle^{c}
_{{\cal S}_{l}}-G_{l}\right):G_{l}^{-1}:G
\end{equation}
and so on for higher correlations. When performing a perturbative
calculation, the factors $G:G_{l}^{-1}$ restore the original
propagator for the external lines, whereas internal lines of the graph involve
$G_l$. Accordingly, with this procedure the function $c(s)$ can be arbitrary
(it is however convenient - see below - to use $c'(0)=0$).

Note that if $G:J=G_{l}:J$, one recovers (\ref{ejphi}), i.e. $W (J)=W_{l}
(J)$ for these slow $J$'s as a special
case of (\ref{examplecutoff}). In that case, for $q<\Lambda '$, (\ref{phiphi})
reduces to $\langle \phi \phi \rangle^{c}_{{\cal S}}=\langle \phi \phi
\rangle^{c}_{{\cal S}_{l}}$ as it should.

To compute correlation functions, it is 
useful to express $W (J)$ in a perturbation expansion in powers
of ${\cal V}_{l} (\phi)$, which reads to lowest order
\begin{eqnarray}\label{wjocalv}
W(J)=\frac{1}{2}J:G:J-e^{-\frac{1}{2}J:G:G_{l}^{-1}:G:J}
\left[e^{J:G:G_{l}^{-1}:\phi}{\cal V}_{l} (\phi ) \right]_{G_{l}}+{\cal O}
({\cal V}_{l} ^{2})
\end{eqnarray}

The Polchinski equation (\ref{poleq}) can equivalently be written as a 
functional ``diffusion'' equation 
\begin{eqnarray}\label{diffusion}
\partial_{l}e^{-{\cal V}_{l}(\phi
)}=-\frac{1}{2}{\rm Tr}\partial_{l}G_{l}:\frac{\delta ^{2}}{\delta
\phi \delta \phi }e^{-{\cal V}_{l}(\phi)}
\end{eqnarray}
or in its integrated form
\begin{eqnarray}
e^{-{\cal V}_{l}(\phi
)}=\left[e^{-{\cal V}_{0}(\phi + \phi' 
)} \right]_{G-G_{l}}
\end{eqnarray}
where the average is over $\phi' $, which makes explicit the
definition of ${\cal V}_{l} (\phi )$ as a coarse-grained interaction,
i.e. integrated over the ``fast part''  $\phi '$ 
of the field. In fact, the decomposition into slow and fast modes and
the definition of coarse-grained observables relies on the property
$\left[ {\cal A} 
(\phi )\right]_{G}=\left[ \left[ {\cal A} (\phi
+\phi')\right]_{G-G_{l}} \right]_{G_{l}}$ of Gaussian averages (see
Appendix~\ref{coarsegrain}).

Although in general the Polchinski equation is far too complicated
to be solved, in some simple cases one can find exact solutions
e.g. Gaussian models, zero dimensional toy model. Most interestingly,
there exist a large class of exact solutions
which appear as superpositions of gaussians. In all these cases, one
can explicitly verify an interesting property of the Polchinski equation
to generate cusp singularities. This is further discussed 
in Appendix~\ref{examples} and in forthcoming publication \cite{us_inprep}.

\subsection{Multi-local expansion}
\label{multiloc}

The Polchinski equation, in addition to being elegant, is conceptually
more satisfactory than other RG methods, e.g. Wilson's shell
renormalization, because it is exact and better
controlled since. Being valid for arbitrary cutoff procedures, 
it does not suffer from the problems 
associated with the sharp cutoff~\cite{morijmp}. However, this
functional equation generates non-local operators, which until now,
has limited its practical applications. This generation 
can be seen in terms of
Feynman diagrams and compared to Wilson's
shell renormalization, since $G-G_{l}$ which contains a range of
wave-vectors centered around 
$\Lambda _{0}e^{-l}$, plays the role of the on-shell
propagator. The term ${\cal V}''$ with a second derivative
in (\ref{poleq}) 
represents tadpoles while the term ${\cal V}'{\cal V}'$ represents diagrams
with only one contraction (one particle reducible). 
These last terms are non-local operators. For
instance in $\phi ^{4}$ theory, it generates the operator $\phi 
(x)^{3} \partial G^{x-y} \phi (y)^{3}$ which is bi-local
since it corresponds to a graph where external momenta must be greater
than $\Lambda _{l}$. The way Polchinski's equation reproduces the
loop diagrams (i.e. local terms) is that after integration over a
slice $dl$, a bilocal interaction generated by the second term of
(\ref{poleq}) is fed into tadpole diagrams. A fast momentum
goes around the corresponding loop, and slow external momenta are
allowed. Thus one needs to integrate the flow and study the feedback of
the generated non-local operators into local ones.

We now present a method which allows to perform
this program in a controlled way. The following
expansion in the number of points (local, bi-local, etc..)
\begin{eqnarray}\label{uv}
{\cal  V} (\phi )=\int_{x}U (\phi^{x})+\int_{xy}V
(\phi^{x}, \phi^{y}, x-y )+\dots 
\end{eqnarray}
is valid a priori for any translationally invariant functional 
${\cal  V} (\phi )$ interactions. We
discuss here only the first two terms, the general systematics being
given in Appendix~\ref{higher}. Here, $U (\phi )$ is a function of
the vector $\phi _{i}$ and involves the value of the field at one
point in space. The bi-local part is a function $V (\phi ,\psi ,z)$ of
two vectors $\phi $, $\psi $ and a space coordinate difference
$z$. In order that the expansion be well-defined, one needs the
bi-local interactions to have {\it no} projection on the local ones. A
natural way to define such a projection, inspired from the conventional
short-distance-expansion, is the exact equality
\begin{eqnarray}
\int_{xy}F (\phi ^{x},\phi ^{y},x-y)&=&\int_{xy} F (\phi ^{x},\phi
^{x},y) + \int_{xy}\left(F (\phi ^{x+y/2},\phi ^{x-y/2},y) -F (\phi ^{x},\phi
^{x},y) \right)\\
&=&\int_{x} (\overline{P}_{1}F) (\phi ^{x})+\int_{xy} ((1-P_{1})F) (\phi ^{x},\phi ^{y},x-y)
\end{eqnarray}
where we have introduced the projections 
\begin{eqnarray}
(\overline{P}_{1}F) (\phi
)&=&\int_{y} F (\phi ,\phi ,y) \label{projo1}\\
(P_{1} F) (\phi,\psi ,z)&=& \delta (z)\int_{y} F (\phi ,\psi ,y)
\label{projo2}
\end{eqnarray}
on the subspaces of local and bi-local interactions
respectively. Indeed, $(\overline{P}_{1} (1-P_1)F )(\phi ) 
=\int_{x} ((1-P_1)F) (\phi ,\phi,x)=0$, i.e. $(1-P_1)F$ has zero local 
part and is thus properly bi-local. Interestingly, this definition
implies that the function $V (\phi ,\psi ,z)$ appearing in the proper
bi-local operator of (\ref{uv}) also satisfy the stronger property
$\int_{z} V (\phi,\psi ,z)=0$ for any $\phi ,\psi $. Note that with no loss of
generality, $V (\phi ,\psi ,z)=V (\psi ,\phi ,-z)$. Here in
addition, we will consider parity invariant theories ($V (\phi ,\psi
,z)=V (\phi ,\psi ,-z)$ too).

For theories where the initial interaction $U$ is local and is formally 
treated as a ``small'' quantity $U$ (e.g. the $\phi^4$ theory in 
$D=4-\epsilon$ where $U \sim \epsilon$), it is natural to
consider that the bi-local term will be of higher order ${\cal O}
(U^{2})$. In fact, this property results from the Polchinski equation
since the term which creates bilocal interactions from local ones is
${\cal O} (U^{2})$ (the first part ${\cal  V}''$ does not increase the
degree of non-locality of ${\cal  V}$). This property that solutions of the
Polchinski equation can be organized in powers of $U$ depending on
their locality holds to arbitrary orders ($p$-local operators
are ${\cal O} (U^{p})$) as is discussed below and
shown in Appendix~\ref{higher}. Thus, to lowest non-trivial order
${\cal O} (U^{2})$, the flow equations involve 
local and bilocal parts. Their schematic structure is
\begin{eqnarray}
\partial U &=& U'' + \overline{P}_{1} (V''+ U'U')+{\cal  O} (UV)+{\cal  O}
(V^{2})\label{paru}\\
\partial V &=& (1-P_1)( V'' + U'U' ) +{\cal  O} (UV)+{\cal  O} (V^{2})
\label{parv}
\end{eqnarray}
where we have written the subdominant terms which will be neglected in
the following. 

A simplification occurs if we choose, as done in this Section,
$(\partial_{l}G_{l})_{ij} (q=0)=0$. Indeed, the term $\overline{P}_{1}U'U'$
vanishes since
\begin{eqnarray}
\int_{x}  (\partial_{l}G^{x}_{l})_{ij}  \partial_{i}U (\phi)
\partial_{j}U (\phi)=0
\end{eqnarray}
Let us write now (\ref{paru},\ref{parv}) in an explicit form:
\begin{eqnarray}
&& \partial_{l} U_{l} (\phi )=-\frac{1}{2}\partial G_{ij}^{x=0}
\partial_{i}\partial_{j}U_{l} (\phi )-\int_{x} \partial G^{x}_{ij}
\partial^{1}_{i}\partial^{2}_{j}V_{l} (\phi ,\phi ,x)\\
&& \partial_{l}V_{l} (\phi,\psi ,x
)=-\partial^{1}_{i}\partial^{2}_{j}\left( \partial G_{ij}^{x} 
V_{l} (\phi ,\psi ,x)-\delta (x)\int_{y}\partial  G_{ij}^{y} V_{l} (\phi
,\psi ,y)\right) \\
&&
-\frac{1}{2}\partial G_{ij}^{x=0}\left(\partial^{1}_{i}\partial^{1}_{j}+
\partial^{2}_{i}\partial^{2}_{j}\right) V_{l} (\phi ,\psi ,x)
+\frac{1}{2}\partial G^{x}_{ij}\partial_{i}U_{l} (\phi )\partial_{j}
U_{l} (\psi )
\end{eqnarray}
where $\partial G$ stands for $\partial_{l}G_{l}$ and
$\partial^{1}_{i}A (\phi ,\psi )$ (resp. $\partial^{2}_{i}A (\phi
,\psi )$) for $\frac{\partial}{\partial \phi_{i} }A (\phi ,\psi )$
(resp. $\frac{\partial}{\partial \psi_{i} }A (\phi ,\psi )$).

\subsection{Solution to the lowest order RG-equations}

\label{sollow}

To solve (\ref{paru},\ref{parv}), one switches to Fourier space (in the
field):
\begin{eqnarray}
U^{K}&=&\int d\phi \, e^{-iK.\phi} U (\phi )\\
V^{KPx}&=&\int d\phi \, d\psi \, 
e^{-iK.\phi-iP.\psi}\, V (\phi,\psi ,x )
\end{eqnarray}
where $K.\phi \equiv \sum_{i} K_{i}\phi _{i}$. It turns out that the
equation for $V_{l}$ can be integrated explicitely as a retarded function of
$U_{l}$:
\begin{eqnarray}
V^{KPx}_{l}&=&\frac{1}{2}\left(F_{l}^{KPx}-\delta
(x)\int_{y}F_{l}^{KPy} \right)\label{vdef}\\
F_{l}^{KPx}&=& -\int_{0}^{l}dl'\, \left( K.\partial G_{l'}^{x}.P\right)
U^{K}_{l'}U^{P}_{l'} e^{\frac{1}{2}K.(G^{x=0}_{l} 
-G^{x=0}_{l'}).K+\frac{1}{2} P.(G^{x=0}_{l} -G^{x=0}_{l'}).P 
+K.(G^{x}_{l} -G^{x}_{l'} ).P}\label{fdel}
\end{eqnarray}
since we have chosen $V^{KPx}_{l=0}=0$.
One can then reinject this result in (\ref{paru}) and obtain a closed
RG equation for $U_{l} (\phi )$:
\begin{eqnarray}
\partial_{l}U_{l} (\phi )&=&-\frac{1}{2}\partial
G^{x=0}_{ij}\partial_{i}\partial_{j}U_{l} (\phi )
-\frac{1}{2}\int_{KP} e^{i(K+P).\phi }
\int_{x}K.(\partial G^{x}_{l}-\partial G^{x=0}_{l}).P\nonumber \\
&&\int_{0}^{l }dl' \,
\, K.\partial G^{x}_{l'}.P\,
e^{\frac{1}{2}K.(G^{x=0}_{l} 
-G^{x=0}_{l'}).K+\frac{1}{2} P.(G^{x=0}_{l} -G^{x=0}_{l'}).P 
+K.(G^{x}_{l} -G^{x}_{l'} ).P} U^{K}_{l'}U^{P}_{l' }\label{exact}
\end{eqnarray}
This is the {\it exact} renormalization equation for an arbitrary
local interaction $U (\phi )$ to ${\cal O} (U^{2})$. Note that the
second order term is retarded in $l$, since as discussed above, local
terms are generated only after integration.

More generally, this procedure can be carried out to any order in $U$
using the hierarchical structure of the flow equations for $p$-local
interactions (see Appendix~\ref{higher}).
It is found that the general structure for the flow of the local part is
\begin{eqnarray}\label{beta}
\partial_{l}U_{l} (\phi) &=& \beta \left[U_{l'<l} (\phi ') \right](\phi )\\
&=& \sum_{n\geq 1} \int_{l_{1}<l_{2}<\dots <l_{n}<l} 
{\cal K}^{n}_{l,l_{1}.. l_{n-1}}[\frac{\partial}{\partial \phi_1},..
\frac{\partial}{\partial \phi_n}]
U_{l_{1}} (\phi_1)\dots U_{l_{n}} (\phi _{n})|_{\phi_{p}=\phi}
\end{eqnarray}
and in (\ref{exact}) we achieved the calculation of the $\beta$
function to second order in $U$.
Once the solution of (\ref{beta}) is known up to ${\cal O} (U^{p})$,
all the $p'$-local flowing interactions for $p'\leq p$ are also known
by injecting the solution for $U_{l}$. For example, for $p=2$, the
bilocal part is obtained from (\ref{vdef},\ref{fdel}) 
injecting the solution to (\ref{exact}).

To a given order in $U$ one can also perform a loop expansion by
expanding the exponentials of propagators which appear in
(\ref{exact}) (see Appendix~\ref{rgu} for the ${\cal O} (U^{3})$
equation). To order $U^2$ and
one loop it reads
\begin{eqnarray}
\partial_{l} U_{l }(\phi) &=& J^{0}_{l} \partial_{ii} U_{l }(\phi)
+ \int_{0}^{l} dl' \, J^D_{l l'} 
\partial_{ij} U_{l'}(\phi) \partial_{ij} U_{l'}(\phi)  
\label{onelooprggeneral} \\
J^0_l &=& - \frac{1}{2} \int_{q} \partial G_l^{q} \\
J^D_{l l'} &=& - \frac{1}{2} \int_{q} \partial G_l^{q} \partial G_{l'}^{q} 
\end{eqnarray}

In order to compute the correlation functions for small wave-vectors $q$, the
strategy of the RG consists in performing a perturbative calculation
in the theory renormalized up to $l'=\ln (\Lambda _{0}/q)$. In the
favorable cases, the interaction ${\cal V}_{l}$ flows, from a small
initial initial interaction ${\cal V}_{0}$, to ``fixed points'' 
in functional space (up to appropriate rescalings) controlled by a
small parameter (such as the offset from the critical dimension). Once
the asymptotic large $l$ behaviour of $U_l(\phi)$ is known, one uses
the invariance property of $W (J)$ (see (\ref{wwj}) and
Appendix~\ref{correl} where this is done in details) to compute the
observables. To lowest order in $U$, it is sufficient 
to keep only the local part in (\ref{wjocalv}) which yields
\begin{eqnarray}
W (J)=\frac{1}{2}J:G:J-\int_{x}\int_{K}U^{K}_{l}
e^{-\frac{1}{2}K.G^{x=0}.K} e^{iK.(G:J)^{x}} +{\cal O} (U^{2}_{l})
\end{eqnarray}
Thus one has for the two-point function $\langle \phi ^{q}_{i}
\phi ^{q'}_{j}\rangle ^{c}_{\cal S} = \delta (q+q') {\sf C}^{q}_{ij}$ with
\begin{eqnarray}
&&{\sf C}^{q}_{ij}=G^{q}_{ij} + \int_{K}(K.G^q )_i
(K.G^q)_j \hat{U}^{K}_{l} \\ 
&&\hat{U}^{K}_{l}=U^{K}_{l}e^{-\frac{1}{2}K.G^{x=0}_{l}.K}
\end{eqnarray}
and more generally the $n$-point function ($n\neq 2$):
\begin{eqnarray}
\langle \phi ^{q_{1}}_{j_{1}}\dots \phi ^{q_{n}}_{j_{n}}\rangle
^{c}_{\cal S}=-\delta_{\sum_{i}q_{i}}\left(\prod
_{i}G^{q_{i}}_{j_{i}k_{i}} \partial_{k_{i}}\right) 
\hat{U}_{l} (0)
\end{eqnarray}
with $\delta _{q}\equiv (2\pi )^{D}\delta (q)$. 
To compute e.g. the two point correlation function at wave-vector $q$, 
one carries perturbation theory in $U_{l'}$ at a large scale $l'$ and
it is convenient to choose $l'=\ln \Lambda/q$.
To first order in $U_{l'} (\phi )$, one has
\begin{eqnarray}\label{sfC}
&&{\sf C}^{q}_{ij}=G^{q}_{ij}  + \int_{K}(K.G^q )_i
(K.G^q)_j U^{K}_{l'=\ln \Lambda /q} e^{-\frac{1}{2}K.G^{x=0}_{l'=\ln
\Lambda /q}.K}  
\end{eqnarray}
Of course, since $W(J)$ is by construction independent of $l$ the result
should not depend on the choice of $l'$. Using the RG flow equation,
it can be checked order by order in perturbation in powers of $U_l$
that this is the case.

In the above computation of the two point
function, the natural vertex which appear is not $U_l(\phi)$ but 
$\hat{U}(\phi) = \int_K e^{i K . \phi} \hat{U}^{K}_{l}$; it is thus interesting
to study directly its flow equation.

\section{Removing of tadpoles and application to $\phi^4$}

\label{removing}

\subsection{Modified Polchinski equation}

\label{modpol}

\begin{figure}[hbt]
\centerline{\epsfig{file=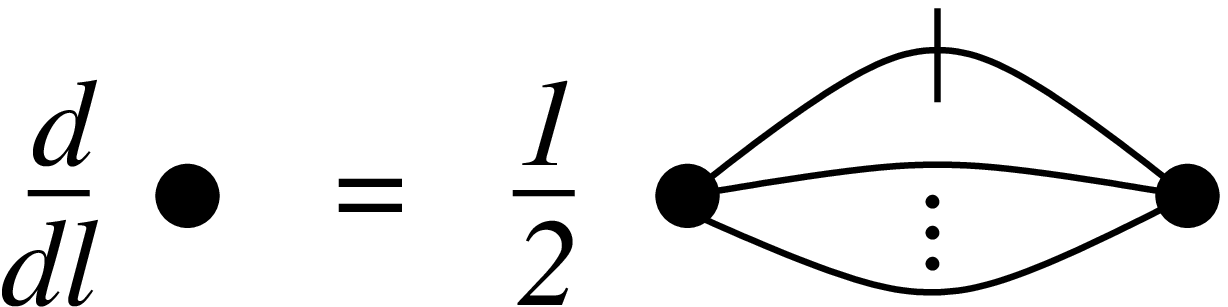,angle=0,width=5.cm}}
\caption{Graphical representation of the modified Polchinski equation.
The point represent any vertex, the broken line is a propagator
on shell $\partial G_l$ and the full line is a $G_l$.}
\label{modified}
\end{figure}

It is useful for some applications, and in particular to simplify
higher orders calculations, to get rid of the linear term in the Polchinski
equation. This can be achieved exactly by introducing the following
functional:
\begin{eqnarray}\label{hatv}
\hat{\cal V}_{l}(\phi)=e^{\frac{1}{2}\frac{\delta}{\delta \phi }:G_l :
\frac{\delta}{\delta \phi }}{\cal V}_{l}(\phi)
\end{eqnarray}
Inserted in (\ref{poleq}) one finds that it satisfies
\begin{eqnarray}\label{polhat}
\partial_{l} \hat{\cal V}_{l}(\phi)=\frac{1}{2} 
e^{\frac{\delta}{\delta \phi _{1}}:G_l :\frac{\delta}{\delta \phi _{2}}} 
\frac{\delta}{\delta \phi _{1}}:\partial G_l :\frac{\delta}{\delta \phi _{2}}
\hat{\cal V}_{l}(\phi_{1})|_{\phi_{1}=\phi}
\hat{\cal V}_{l}
(\phi_{2})|_{\phi_{2}=\phi}
\end{eqnarray}
The graphical representation of this equation is drawn in Fig.~\ref{modified}. 
Since this equation does not contain a linear term,
its solution does not contain tadpole-like diagrams. This functional 
has thus several advantages: first it enters directly the computation 
of any observable, second its flow is 
simpler than the one of the bare vertices. Finally in the context of 
quantum field theory it has the direct meaning of being the normally ordered 
vertices.

We now perform a multilocal expansion, similar to the one introduced
in the previous Section, but on the functional
$\hat{\cal V}_{l}(\phi)$ as:
\begin{eqnarray}
\hat{\cal V}_{l}(\phi) = \int_x \hat{U}_{l}(\phi^x) 
+ \int_{xy} \hat{V}_{l}(\phi^x,\phi^y,x-y) + ..
\end{eqnarray}
The modified Polchinski equation (\ref{polhat}) can be solved
order by order in $\hat{U}_l(\phi)$. The general analysis is performed in 
the Appendix~\ref{vhatmulti}. Here we give only the result to
order $\hat{U}^2$ which reads:
\begin{eqnarray}
\partial_{l} \hat{U}_l(\phi)&=&\frac{1}{2} \int_{x} 
e^{\partial^{1} G^x_l \partial^{2}}
\partial^{1} \partial G^x_l \partial^{2}
\hat{U}_l(\phi_{1}) |_{\phi_{1}=\phi}
\hat{U}_l(\phi_{2}) |_{\phi_{2}=\phi}
\label{betahat1}
\end{eqnarray}
An interesting property of this equation is that it is now local in
$l$. 

Expansion in the number of loops $k$, restricted 
to order $\hat{U}_l^2$, is thus straightforward:
\begin{eqnarray}
&& \partial_{l} \hat{U}_l(\phi) = \frac{1}{2} 
\sum_{k=1}^{+\infty} \frac{1}{k!} I^k_l 
\partial_{i_1,..i_{k+1}} \hat{U}_l(\phi)
\partial_{i_1,..i_{k+1}} \hat{U}_l(\phi) \\
&& I^k_l = \int_x (G_l^x)^k \partial G^x_l = \frac{1}{k+1}
\partial_l \int_x (G_l^x)^{k+1}
\end{eqnarray}

Note that the multilocal expansion for $\hat{\cal V}$ and ${\cal V}$ 
are not identical,
i.e they do not produce order by order equations which can be transformed 
back into each others. However, they should yield the same result at the end 
when calculating universal fixed point quantities. This property
will be tested and used in the rest of the paper. 

We also give the expression of the bilocal term, as one must check that
it effectively reaches a fixed point for consistency. It reads
as a function of the local term:

\begin{eqnarray}
\hat{V}_l(\phi_{1},\phi_2,q) =
\frac{1}{2} \int_x (e^{i q x} -1) \int_0^l dl'
\, e^{\partial^{1} G^x_{l'} \partial^{2}}
\partial^{1}\partial G^x_{l'}\partial^{2}  
\hat{U}_{l'}(\phi_{1}) 
\hat{U}_{l'}(\phi_{2})
\end{eqnarray}

\subsection{Modified RG equation to one loop 
and application to $O(n)$ model}
\label{on}

Expanding the exponential of the propagator in (\ref{betahat1})
yields a loop expansion of the beta function to order $U^2$.
To one loop this gives:
\begin{eqnarray}\label{polhat31}
&& \partial_{l} \hat{U}_l (\phi) = I_l^D 
\partial_{ij} \hat{U}_{l}(\phi) \partial_{ij} \hat{U}_{l}(\phi) \\
&& I_l^D = \frac{1}{2} \int_{q} \partial G_{l}^q G_{l}^q
\end{eqnarray}

As a simple application let us consider the $O(n)$ model in
$D=4-\epsilon$ with the polynomial interaction:

\begin{eqnarray}
U_l(\phi) = \frac{1}{2} g_{2 ,l} \phi^2 
+ \frac{1}{4!} g_{4,l} (\phi^2)^2 + \frac{1}{6!} g_{6 ,l} (\phi^2)^3 + ..
\end{eqnarray}
the dimensionless variables being $\tilde{g}_{n,l}=g_{n,l} 
\Lambda_l^{\frac{D-2}{2} n - D}$. The RG can be performed either using 
$U_l(\phi)$ as in the previous section, or in terms of $\hat{U}_l (\phi)$,
by which we start. One has:

\begin{eqnarray}
\hat{U}_l(\phi) = e^{\frac{1}{2} G_l^{x=0} \nabla^2} U_l(\phi)
= \frac{1}{2} a_{2,l} \phi^2 
+ \frac{1}{4!} a_{4 ,l} (\phi^2)^2 + \frac{1}{6!} a_{6, l} (\phi^2)^3 + ..
\end{eqnarray}
and, similarly~\cite{footnote3} $\tilde{a}_{n,l}=a_{n,l}
\Lambda_l^{\frac{D-2}{2} n - D}$ thus:
 
\begin{eqnarray}
&& \tilde{a}_{2,l} = a_{2,l} \Lambda_l^{- 2} \\
&& \tilde{a}_{4,l} = a_{4,l} \Lambda_l^{- \epsilon} \\
&& \tilde{a}_{6,l} = a_{6,l} \Lambda_l^{2 - 2 \epsilon} 
\end{eqnarray}

From $
\partial_{ij} \hat{U}_{l}(\phi) = a_{2,l} \delta_{ij} +
\frac{a_{4,l}}{3!} (\delta_{ij} \phi^2 + 2 \phi_i \phi_j) 
+ \frac{a_{6,l}}{5!} (\delta_{ij} (\phi^2)^2 + 4 \phi_i \phi_j \phi^2) + ..
$, one obtains the RG equations:
\begin{eqnarray}
&& \partial \tilde{a}_{2 ,l} = 2 \tilde{a}_{2 ,l} 
+ I^D \frac{2}{3} (n+2) \tilde{a}_{4, l} \tilde{a}_{2,l} + {\cal O}
(\tilde{a}_{4 ,l}^2)
\\
&& \partial \tilde{a}_{4 ,l} = \epsilon \tilde{a}_{4, l} + I^D \frac{2}{3} (n+8)
\tilde{a}_{4,l}^2 + {\cal O}(\tilde{a}_{4 ,l}^3) \\
&& \partial \tilde{a}_{6  ,l} = (-2 + 2 \epsilon) 
\tilde{a}_{6, l} + {\cal O}(\tilde{a}_{4, l}^3)
\end{eqnarray}
in terms of the single integral $I^D$ which has a 
universal value in $D=4$:
\begin{eqnarray}
&& I^D = \Lambda_l^\epsilon I^D_l = \Lambda_l^\epsilon \frac{1}{2} 
\int_{q} \frac{c'(\frac{q^2}{2 \Lambda_l})}{\Lambda_l^2}
\frac{c(\frac{q^2}{2 \Lambda_l})}{q^2} \\
&& = \frac{1}{2}  S_D \int_{s>0} (2 s)^{- \epsilon/2} c'(s) c(s) 
= - \frac{S_4}{4} + {\cal O}(\epsilon)
\end{eqnarray}
with $S_4=1/(8 \pi^2)$ and here and in the following $S_D$ is the unit sphere
area divided by $(2 \pi)^D$. Thus $\tilde{a}_{4 ,l}$ flows to the
fixed value:
\begin{eqnarray}
\tilde{a}_{4}^* = \frac{3}{(n+8)} 16 \pi^2 \epsilon
\end{eqnarray}
which is universal to this order.
We have indicated terms in the above RG equations which arise at two loops 
and yield fixed point for $\tilde{a}_{2 ,l}$ and $\tilde{a}_{6 , l}$ with:
\begin{eqnarray}
\tilde{a}_{2}^* = {\cal O}(\epsilon^2) \quad , \quad
\tilde{a}_{6}^* = {\cal O}(\epsilon^3)
\end{eqnarray}
and more generally $\tilde{a}_{2 n}^* = {\cal O}(\epsilon^{n})$ for $n \ge
3$. The derivation
of this is simple but goes beyond this paper~\cite{us_part2}. 
This fixed point is unstable in the direction
$\tilde{a}_{2}$ and stable in all other directions. The stability
eigenvalues to ${\cal O}(\epsilon)$ read:
\begin{eqnarray}
&& \lambda_2 = 2 - \frac{n+2}{n+8} \epsilon \\
&& \lambda_4 = - \epsilon  \\
&& \lambda_{2 n} = - 2(n-2) + \epsilon (n-1) \quad n \ge 3
\end{eqnarray}

The critical manifold of the $O(n)$ model corresponds to 
$\tilde{a}_{2} = \tilde{a}_{2}^*$. This corresponds indeed
to the massless case, since the self energy
\begin{eqnarray}
\Sigma^q = \hat{U}''(0) + {\cal O}(\hat{U}^2) = \Lambda_l^2 \tilde{a}_{2,l}
\end{eqnarray}
vanishes to lowest order on the critical manifold. The instability eigenvalue 
at the the fixed point gives the critical exponent $\gamma$:
\begin{eqnarray}
\gamma = \frac{2}{\lambda_2} = 1 + \frac{(n+2)}{2 (n+8)} \epsilon +
{\cal O} (\epsilon ^{2})
\end{eqnarray}
thus recovering the standard result. One also gets $\omega=- \epsilon$
and $\eta={\cal O}(\epsilon^2)$. 

Note that to first order in $\epsilon$ there is no $q$ dependence 
of $\Sigma^q$ and no wave function renormalization. This can be incorporated
in the method to two loops (see Appendix~\ref{zphi} and \cite{us_part2}).

\section{Application to disordered elastic systems}

\label{elastic}

\subsection{FRG equations and universality to one loop}

\label{frgequniv}

\subsubsection{The model}

Let us consider an elastic system of internal dimension
$D$ embedded in a disordered medium. It is described by a single
component displacement field $u^{x}$ ($x$ is the internal variable), 
which is either the height function for an interface problem, or the
continuous 
deformation field for periodic systems. The energy reads
\begin{eqnarray}
H (u) =  \int_{x} \left(\frac{c}{2}|\nabla u^{x}|^{2} -W (x,u^{x}) 
+ \frac{m^2}{2} |u^{x}|^2 \right)
\label{interface}
\end{eqnarray}
where a short-distance cutoff is implicit. The elastic constant is set
to $c=1$ here, and the mass to $m=0$, its effect will be studied
in Section~\ref{biased}. 
The disordered potential $-W (x,u)$ is a random variable which has the
following properties (i) $\overline{W (x,u)}=0$ (ii) the potential at
different $x$ are uncorrelated (iii) the distribution of $W (x,u)$ is
translationally invariant in $u$ space. Its cumulants read
\begin{eqnarray}\label{S(N)}
\overline{W (x_{1},u_{1})\dots W
(x_{N},u_{N})}^{c}=\delta (x_{1}-x_{2})\dots \delta (x_{1}-x_{N})
S^{(N)} (u_{1},\dots ,u_{N})
\end{eqnarray}
where the symmetric
functions $S^{(N)}$ satisfy $S^{(N)}
(u_{1},\dots ,u_{N})=S^{(N)} (u_{1}+u,\dots ,u_{N}+u)$ for any $u$. 
In particular,
the second cumulant is denoted by $R (u-u')=S^{(2)} (u,u')$. In the
case of an interface, these correlators can be either long--range,
e.g. random field, or short--range, e.g. random bond. In the periodic
case, the cumulants have the periodicity of the lattice. We assume
parity symmetry $S^{(N)} (-u_{1},\dots ,-u_{N})=S^{(N)} (u_{1},\dots ,u_{N})$.
This problem is usually studied by introducing $n$ replicas
$\phi _{a}^{x}$, $a=1\dots  n$, of $u^{x}$ and by averaging over the
disorder. It yields the action
\begin{eqnarray}\label{relicaaction}
{\cal S}(\phi )=\int_{x}\left[\frac{1}{2T}\sum_{a}|\nabla \phi_{a }^{x}
|^{2}-\sum_{N\geq 2}\frac{1}{N!T^{N}}\sum_{a_{1}\dots a_{N}}
S^{ (N)} (\phi ^{x}_{a_{1}}, \dots , \phi ^{x}_{a_{N}}) \right]
\end{eqnarray}
Thus the bare ${\cal S}$ has the general form (\ref{generalaction}) with
\begin{eqnarray}
&&G^{q}_{ab}=\frac{T}{q^{2}}\, c (\frac{q^{2}}{2\Lambda^{2}})\, \delta _{ab}\\
&&{\cal V} (\phi )=\int_{x}U (\phi ^{x})\\
&&U (\phi )= -\frac{1}{2T^{2}}\sum_{ab}R (\phi
_{a}-\phi _{b}) - \frac{1}{3!T^{3}}\sum_{abc}S^{(3)} (\phi
_{a},\phi _{b}, \phi _{c}) +\dots \label{rsN}
\end{eqnarray}
We will consider any cutoff function $c (s)$ such that $c (0)=1$ with
no loss of generality and with 
$c' (0)=0$ for convenience (see below).

Direct perturbation theory can be performed on this model. One can show that
it has a well defined $T=0$ limit (see Appendix \ref{fleurs}). Furthermore,
at $T=0$ this perturbation theory is in fact trivial, i.e the disorder average
of any observable is identical to its value in the linear random force
model, as shown in the Appendix \ref{fleurs}. Within the exact RG one can
in fact escape this well known dimensional reduction phenomenon, since,
as we will see below, the flowing disorder becomes non analytic. As shown below
it yields non trivial results for correlations.

\subsubsection{RG analysis}

We now use the exact RG method introduced above. For now we use the
RG equations based on the multilocal expansion of ${\cal V}$ 
while
the other method in terms of $\hat{\cal V}$ (explained in Section~\ref{modpol} 
and Appendix~\ref{vhatmulti}) will be used in Section~\ref{biased}. 
The method
with ${\cal V}$ turns out to be more convenient to analyze finite $T$
effects. The
$l$ dependence of the Gaussian part is implemented by the choice
\begin{eqnarray}
( G_{l} )^{q}_{ab}= T ( \overline{G}_{l} )^{q}_{ab}
=\frac{T}{q^{2}}\, c (\frac{q^{2}}{2\Lambda_{l}^{2}})\, \delta _{ab}
\end{eqnarray}
where $\Lambda _{l}=\Lambda e^{-l}$.
This choice is particularly convenient here since there is no
correction to any order to the connected quadratic part (statistical
tilt symmetry\cite{brezin_sts}).
The flowing interaction functional ${\cal V}_{l} (\phi )$ 
remains translationally and parity
invariant in $x$ space. 
Since translation invariance in the $u$ space is conserved, its local
part $U_{l} (\phi )$ remains of the form  (\ref{rsN}). In order to
obtain fixed points it is convenient to define a rescaled dimensionless
temperature $\tilde{T}_{l}=T\Lambda _{l}^{D-2}$ and rescaled functions
\begin{eqnarray}
\tilde{U} _{l} (\phi ) &=& \Lambda _{l}^{-D}U_{l} (\phi )\\
&=&-\frac{1}{2\tilde{T}_{l}^{2}}\sum_{ab}\tilde{R}_{l} (\phi
_{a}-\phi _{b}) - \frac{1}{3! \tilde{T}_{l}^{3}}\sum_{abc}\tilde{S}^{(3)} (\phi
_{a},\phi _{b}, \phi _{c}) +\dots \\
\tilde{R}_{l} (u )&=&\Lambda _{l}^{D-4}R_{l} (u)\\
\tilde{S}^{(N)}_{l} (u_{1},\dots ,u_{N})
&=&\Lambda_{l}^{DN-2N-D} S^{(N)}_{l}(u_{1},\dots ,u_{N}) 
\end{eqnarray}

The general RG equation (\ref{beta}) for $U_{l} (\phi )$ implies a set
of flow equations for the rescaled cumulants $\tilde{R}_{l} (u)$, 
$\tilde{S}^{(N)}_{l} (u_{1},
\dots , u_{N})$ (since the former is in fact a {\it set} of equations 
for functions of a $n$-dimensional vector $\phi $ for any $n$). 
The rescalings above have
been chosen such that these rescaled functions flow to fixed functions
denoted $\tilde{R}^{*} (u)$, $\tilde{S}^{(N)*} (u_{1},\dots , u_{N})$ independent of $T$.

An important property of the theory (\ref{relicaaction}) is that it
admits a well-defined $T=0$ limit, at least at the perturbative level.
This can be seen either by
examination of the diagrammatics (all negative powers of $T$ in the
perturbative calculation of observables are in factor of a positive
power of $n$, see Appendix~\ref{fleurs}), or in the $T=0$ dynamics
\cite{kay_dynamics}. 
Similarly there exists
a well-defined $T=0$ limit of the set of flow equations for the
cumulants. For small $\epsilon =4-D$, this complicated set of coupled
equations can be organized in powers of $\epsilon $. Specifically one
finds that $\tilde{R}^{*}={\cal O} (\epsilon )$ while $\tilde{S}^{(N)*}={\cal O}
(\epsilon ^{N})$ for $N\geq 3$. This can be seen on the schematic
structure (ordered in $U$ and $T$) 
of the flow equations obeyed by the rescaled
$\tilde{R}_{l}$, $\tilde{S}^{(N)}_{l}$, which can be read off
from Appendix~\ref{rgu}:
\begin{eqnarray}
\partial U = T U +  T^2  U^2 e^T + T^3 U^3 e^T +\dots 
\end{eqnarray}
The two first terms reproduce (\ref{exact}) since $\partial G\propto
T$, while the third term mimics the ${\cal O} (U^{3})$ in the $\beta $
function. Its three $U$ vertices must be linked by at least three
propagators because of the constraint of locality. Substituting symbolically
$\tilde{U}=\tilde{R}/T^{2}+\tilde{S}/T^{3}$, where we restrict to the
two lowest cumulants, one finds 
\begin{eqnarray}
&& \partial \tilde{R} = \epsilon \tilde{R} + \tilde{S}|_2 + \tilde{R}^2|_2 + 
\tilde{R} \tilde{S}e^T/T|_2 + \tilde{R}^3e^T /T |_2  \\
&& \partial \tilde{S} = (6-2D)  \tilde{S} + \tilde{R} \tilde{S}|_3 + 
\tilde{R}^3|_3 +
\tilde{S}^2 e^T  /T  |_3  + \tilde{R}^2 \tilde{S}e^T /T |_3
\end{eqnarray}
where we have denoted projections on $2$ and $3$ replica parts.
All terms containing $1/T$ vanish after these projections since a
well-defined $T=0$ limit exists.
We have discarded terms, such as $\partial_l \tilde{S} = T \tilde{R}^2|_3$,
which (formally) vanish at $T=0$.

One sees immediately on these equations that the fixed $\tilde{R}^{*}={\cal O}
(\epsilon)$ while $\tilde{S}^{*}={\cal  O} (\epsilon ^{3})$. This can be
generalized by noting that the lowest order (in $\epsilon $)
correction to $S^{(N)}$ is of the form $\tilde{R}^{N}|_{N}$ thus
$\tilde{S}^{(N)*}={\cal O}(\epsilon^{N})$. 
To ${\cal O}(\epsilon^2)$ at $T=0$ we thus need:
\begin{eqnarray}
&& \partial \tilde{R} = \epsilon \tilde{R} + \tilde{R}^2|_2 + \tilde{S}|_2 +
\tilde{R}^3e^T /T  |_2   \\
&& \partial \tilde{S} = (6-2D)  \tilde{S} +  \tilde{R}^3|_3 
\end{eqnarray}
In this paper we simply perform the ${\cal O} (\epsilon )$ calculation
to which we now turn, for which consideration of two replica terms is
sufficient.

We perform the analysis in the $T=0$ limit as explained above.
The propagator can be expressed in terms of
dimensionless quantities as
$G^{x}_{l}=\tilde{T}_{l}
\int_{q}\frac{c(\frac{q^{2}}{2})}{q^{2}}e^{iq \Lambda _{l }x}$. 
At finite $T$, the exponentials of propagators in 
(\ref{exact}) would reduce to $1$ asymptotically at large $l$. This is also
true in the $T=0$ limit for any $l$. It is thus a priori unnecessary
to include higher number of loops within order $U^2$. 

Denoting by $\tilde{M}_{l}(\phi)$ the two--replica term contained in the local
operator $U$ 
\begin{eqnarray}
\tilde{M}_{l}(\phi) = -\frac{1}{2}\sum_{ab} \tilde{R}_{l}(\phi_{a}-\phi_{b})
\end{eqnarray}
the flow equation to one loop (\ref{onelooprggeneral}) (using the change of 
variables $l'\to l-l'$ yields 
\begin{eqnarray}\label{flow1}
\partial_{l} \tilde{M}_{l}(\phi)=(4-D)\tilde{M}_{l}(\phi )
- \frac{1}{2} \int_{0}^{l }dl' \, {\sf K}_{l'}
\sum_{a b }[\partial_{a b }
\tilde{M}_{l -l' }(\phi )]^{2}|_{2-rep}
\end{eqnarray}
where the kernel responsible for the retarded nature of the flow is
\begin{eqnarray}\label{k}
{\sf K}_{l'} = 4 J^D_{l,l-l'} \Lambda_{l-l'}^{2 \epsilon} \Lambda_{l}^{- \epsilon}
= 2 e^{-(6-D)l'} \int_{q} c' (\frac{q^{2}}{2}) 
c'(\frac{q^{2}}{2} e^{- 2 l'})
\end{eqnarray}
Since $c' (u)$ is typically peaked around $u\sim 1$ and decreases fast
at infinity, one sees on (\ref{k}) that the 
range of the kernel ${\sf K}_{l}$ is also of order one and can be made
as small as desired by choosing narrow enough cutoff functions.
The above RG equation (\ref{flow1}) involves computing the contraction:
\begin{eqnarray}
\sum_{ab}\left[\partial_{a}\partial_{b}\tilde{M} (\phi )
\right]^{2}&=&\sum_{ab}\left[\tilde{R}'' (\phi _{a}-\phi_{b}
)^{2}-2\tilde{R}'' (0)\tilde{R}'' (\phi _{a}-\phi_{b} )  \right]\\
&&+\sum_{abc}\tilde{R}'' (\phi _{a}-\phi_{c})\tilde{R}'' (\phi _{c}-\phi_{b})
\end{eqnarray}
where we have used $\partial_{a}\partial_{b}\tilde{M} (\phi )=\delta
_{ab}\sum_{c}\tilde{R}'' (\phi _{a}-\phi _{c})-\tilde{R}'' (\phi_{a}-\phi _{b})$.
The last sum being a three replica term, it does not enter the equation for
$\tilde{R}$ (it is a correction to $S$ proportional to $T$), which reads:
\begin{eqnarray}
\partial_{l}\tilde{R}_{l} (u)=\epsilon \tilde{R}_{l}
(u)+ \int_{0}^{l}dl'\,{\sf K}_{l'} \left( 
\frac{1}{2}\tilde{R}^{\prime \prime }_{l-l'}(u)^2 -
\tilde{R}^{\prime \prime}_{l-l'}(0)  \tilde{R}^{\prime
\prime}_{l-l'}(u)\right) 
\label{frgnonlocal}
\end{eqnarray}

Let us first study the case of periodic elastic systems,
with $\tilde{R}_{l} (\phi )$ periodic of period $1$.
Taking the large $l$ limit we find the fixed point equation:
\begin{eqnarray}
0 =\epsilon \tilde{R}^*(u)+ (\int_{0}^{+\infty} dl'\,{\sf K}_{l'})
\left( \frac{1}{2}\tilde{R}^{* \prime \prime }(u)^2 -
\tilde{R}^{* \prime \prime}(0)  \tilde{R}^{* \prime
\prime}(u)\right) \label{FRG}
\end{eqnarray}
It is now easy to see that the factor ${\sf K}= \int_{0}^{\infty
}dl\, {\sf K}_{l}$ in (\ref{FRG}), which a priori
depends on the dimension of space and of the whole 
arbitrary cutoff function $c (s)$, becomes universal in $D=4$.
Indeed:
\begin{eqnarray}\label{univers}
{\sf K} &=& 2 S_{D}2^{-\epsilon /2} \int_{0}^{\infty}ds\, s^{-\epsilon }c'
(s) \int_{0}^{s} dt\, t^{\epsilon} c' (t)\\
&=& 2 S_{4}\int_{0}^{\infty }ds\, c' (s) (c (s) -1) +{\cal O}(\epsilon ) =
S _{4} +{\cal O}(\epsilon )
\end{eqnarray}
where we used the new variables 
$s=q^{2}/2$, then $t=se^{-2l}$, kept only the lowest order in
$\epsilon$, and used $c (0)=1$. We denote by $S_{D}$ the surface
of the unit sphere in $D$ dimensions divided by $(2\pi )^{D}$.
Thus, to one loop, the FRG equation {\it does not depend on the cutoff 
procedure}. It coincides with the fixed point equation 
obtained\cite{fisher_functional_rg,balents_fisher} from Wilson's 
momentum shell renormalization. 

In Appendix~\ref{loopexpansion}, we also mention the result of a
two-loop calculation of the beta-function in our exact renormalization
framework.

The solution to (\ref{FRG}) is known to be the $1$--periodic
function defined by
\begin{eqnarray}
\tilde{R}^{*} (u) =\frac{\epsilon}{72 S_{4}} \left(\frac{1}{36}-u^{2}
(1-u)^{2} \right) 
\end{eqnarray}
for $0<u<1$. This fixed point function is
non--analytic which is an important and unusual feature. It was argued
in \cite{fisher_functional_rg} 
that this non--analyticity appears at a finite
scale. This scale $R_{c}=e^{l_{c}}/\Lambda $ 
can be identified with the Larkin length at
which metastability and glassiness appears.
Taking the fourth derivative at $u=0$ of (\ref{frgnonlocal}) yields 
a closed retarded equation for $\tilde{R}_{l}^{iv} (0)$ 
\begin{eqnarray}
\partial_{l} \tilde{R}^{iv}_{l} (0) = \epsilon \tilde{R}^{iv}_{l} (0) + 3
\int_{0}^{l}dl'\,{\sf K}_{l'}
\tilde{R}^{iv}_{l-l'} (0)^{2}
\end{eqnarray}
In the limit of narrow cutoffs, the equation becomes local and
$\tilde{R}^{iv}_{l} (0)$ diverges at a finite scale. One can show that
this feature persits in the non--local equation.

The case of an interface (i.e a directed polymer for $D=1$) corresponds to
another fixed point where one must rescale the function $\tilde{R}_l(u)$ 
as follows:
\begin{eqnarray}
\tilde{R}_l(u) = e^{4 \zeta l} r_l(u e^{-\zeta l})
\label{rescalingr}
\end{eqnarray}
and we must now determine $\zeta = {\cal O}(\epsilon)$ such that $r_l(v)$ converges
to a fixed point $r^*(v)$. Inserting (\ref{rescalingr}) into (\ref{frgnonlocal})
yields:
\begin{eqnarray}
\partial_{l}r_{l} (v)=(\epsilon - 4 \zeta) r_{l}(v)
+ \zeta v r_{l}'(v)
+ \int_{0}^{l}dl'\,{\sf K}_{l'} e^{- 4 \zeta l'} \left( 
\frac{1}{2}r^{\prime \prime }_{l-l'}(v e^{\zeta l'})^2 -
r^{\prime \prime}_{l-l'}(0)  r^{\prime
\prime}_{l-l'}(v e^{\zeta l'})\right) 
\label{frgnonlocal2}
\end{eqnarray}
Although the kernel has been modified, this does not affect the
results for the fixed point to lowest order in $\epsilon$. The fixed point 
equation reads:
\begin{eqnarray}
0 =(\epsilon - 4 \zeta) r^*(v)+ \zeta v r^{* \prime}(v)  +
S_4 \left( \frac{1}{2}r^{* \prime \prime }(v)^2 -
r^{* \prime \prime}(0)  r^{* \prime \prime}(v)\right) \label{FRG2}
\end{eqnarray}
and is thus universal, independent of $c(s)$. This shows that $\zeta$,
which, as shown below is the roughness exponent, is universal to one loop.
It will be studied below for the random field case and in the case of
short range disorder it is thus equal to
${\cal O}(\epsilon)$ to the values given in \cite{fisher_functional_rg}.

\subsubsection{Correlation function}

Let us now compute 
the two--point correlation function at $T=0$ using (\ref{phiphi}).
To lowest order in $\epsilon $, it is
sufficient to use the first order formula (\ref{sfC}). The bare
Gaussian part $G^q$ vanishes at $T=0$. We thus get:
\begin{eqnarray}\label{fixedpoint}
{\sf C}_{ab}^{q}=-\frac{T^{2}}{q^{4}}c (\frac{q^{2}}{2\Lambda^{2}}
)^{2}\partial_{a}\partial_{b} U _{l}(\phi )|_{\phi \equiv 0}
=-\frac{R_{l=\ln(q/\Lambda)}^{\prime \prime}
(0)}{q^{4}} = -\frac{\tilde{R}^{* \prime \prime}
(0)}{q^{D}}= \frac{\epsilon}{S_4} \frac{1}{36} q^{-D} 
\end{eqnarray}
where we used that $\tilde{R}_l$ converges to the fixed point 
$\tilde{R}^*$ and small $q$ such that $c(q^{2}/2\Lambda^{2})=1$.
In real space, it yields logarithmic growth of the displacements with a
universal prefactor
\begin{eqnarray}
\overline{\langle (u^{x}-u^{0})^{2}\rangle }&=&
\frac{\epsilon}{18} \ln |\Lambda x|
\end{eqnarray}

In the case of short range disorder (e.g. random bond an for Ising interface)
one gets instead:
\begin{eqnarray}\label{fixedpoint2}
{\sf C}_{ab}^{q}=-\frac{R_{l=\ln(q/\Lambda)}^{\prime \prime}
(0)}{q^{4}} = - e^{2 \zeta l} \Lambda_l^\epsilon
\frac{r^{* \prime \prime} (0)}{q^{D}} \sim
\frac{1}{q^{D + 2 \zeta}}
\end{eqnarray}
This yields to a roughness exponent 
$\overline{\langle (u^{x}-u^{0})^{2}\rangle } \sim |x|^{2 \zeta}$ with
a nonuniversal amplitude (since the FRG fixed point equation
(\ref{FRG2}) is invariant under $r^*(v) \to \lambda^4 r^*(v/\lambda)$
and, contrarily to the periodic case, nothing here fixed the scale).

We can now investigate in more details the structure of the
asymptotic flow of the various higher order interactions 
(three replica terms and higher, as well as bilocal interaction
and more). Although this is beyond the scope of this paper,
such an analysis is in principle necessary for consistency,
i.e to ensure the existence of a global fixed point (for all interactions) 
and the validity of the result to ${\cal O}(\epsilon)$.
We sketch it here for the periodic case $\zeta=0$, generalizations to
interfaces being simple.

We start with estimating the higher cumulants of the renormalized disorder,
i.e the higher replica components of the local interaction $U_l$. 
To lowest order in $R$, the correction to $S^{(N)}_{l}$ is
proportional to $R_{l}^{N}$ and takes the schematic form:
\begin{eqnarray}
\partial \tilde{S}^{(N)}=-(2N+D-DN)\tilde{S}^{(N)} 
+ \tilde{R}^{\prime \prime N} 
\end{eqnarray}
Graphically, the diagram is made of one loop. We dropped the numerous 
higher order terms in $\epsilon$ coming from contractions of various
other cumulants than $R$. One finds that the fixed point 
$\tilde{S}^{(N)*}$ takes the following form
to lowest order in $\epsilon$ 
\begin{eqnarray}
\sum_{a_{1}\dots a_{N}}\tilde{S}^{(N)*} (\phi _{a_{1}}\dots \phi _{a_{N}})=
c_{N,D}\left( {\rm Tr} \left( W^{N}\right)  -
\sum_{a_{1}\dots a_{N},b}\tilde{R}^{*\prime \prime }_{ba_{1}}\dots
\tilde{R}^{*\prime \prime }_{ba_{N}} \right)
\end{eqnarray}
where $W_{ab}(\phi )=\delta
_{ab}\sum_{c}\tilde{R}^{*\prime \prime }_{ac}-\tilde{R}^{*\prime
\prime }_{ab}$, 
$\tilde{R}^{*\prime \prime }_{ab}$ denotes $\tilde{R}^{*\prime \prime
} (\phi _{a}-\phi _{b})$, and $c_{N,D}$ is some number depending on the 
cutoff procedure. The last term in the trace has been
substracted since the product of the $N$ $\delta $'s is a $N+1$ replica term. 
For instance, the third cumulant is of order ${\cal O}(\epsilon^3)$ and
reads
\begin{eqnarray}\label{3rep}
&& \tilde{S}^{(3)*} (u_{1},u_{2},u_{3}) 
=  c_{3,D} \mbox{Sym}_{u_1,u_2,u_3} [ \tilde{R}^{*\prime \prime }
(u_{2}-u_{3})\tilde{R}^{*\prime \prime }
(u_{3}-u_{1})\tilde{R}^{*\prime \prime }(u_{1}-u_{2})\\
&&-3 \tilde{R}^{*\prime \prime }
(u_{1}-u_{2})\tilde{R}^{*\prime \prime } (u_{1}-u_{3})^{2} 
+3 \tilde{R}^{*\prime \prime } (0)\tilde{R}^{*\prime \prime }
(u_{2}-u_{1})\tilde{R}^{*\prime \prime }(u_{3}-u_{1}) ]
\end{eqnarray}
where $c_{3,4}$ is computed in \cite{us_part2} and reads:
\begin{eqnarray}
c_{3,4} = \frac{S_4}{12} \int_0^\infty ds \, \frac{(1-c(s))^3}{s^2}
\end{eqnarray}

Now we check that the bilocal part has a well defined
fixed point. Its expression is given by (\ref{vdef}), where
at $T=0$, the exponentials should be expanded at most to first 
order. The zero-th order term yields three replica terms, while the
first order term yields two replica terms (as well as a correction
proportional to $T$ to three replica terms which we can discard at $T=0$).
Thus we get
\begin{eqnarray}
&& V_l(\phi_1,\phi_2,q) = - \frac{\Lambda_l^{\epsilon}}{2 T^2}
\sum_{ab} \tilde{V}^{(2,2)}_{l}(\frac{q}{\Lambda_l})
- \frac{\Lambda_l^{2\epsilon -2 }}{6 T^3}
\sum_{abc} \tilde{V}^{(2,3)}_{l}(\frac{q}{\Lambda_l})
\end{eqnarray}
where we have explicitly separated two and three replica terms 
respectively:
\begin{eqnarray}
\tilde{V}^{(2,2)}_{l}(\tilde{q}) &=& \frac{1}{2}
\int_k ( \frac{1}{k^2 (k+\tilde{q})^2} (c(k^2/2)-1) (c((k+\tilde{q})^2/2)-1)
- \mbox{idem q=0} ) \nonumber \\
&&(\sum_{ab} \tilde{R}''^1_{ab} \tilde{R}''^2_{ab} \nonumber
 - \tilde{R}''(0) \sum_{ab} (\tilde{R}''^1_{ab} + \tilde{R}''^2_{ab})
) \nonumber\\
&& - \frac{1}{4}  (\int_k  \frac{1}{k^2} c(k^2/2))
\int_0^1 d \alpha (\frac{1}{\alpha}-1) c'(\alpha \tilde{q}^2/2)
(2 \sum_{abc, a \neq b a\neq c} \tilde{R}'''^1_{ab} \tilde{R}'^2_{ac}
+ \tilde{R}'^1_{ab}
\tilde{R}'''^2_{ac})|_{2 rep} \nonumber \\
\tilde{V}^{(2,3)}_{l}(\tilde{q}) &=&
- 3 \frac{c(\tilde{q}^2/2) -1}{\tilde{q}^2} 
\sum_{abc} \tilde{R}'^1_{ab} \tilde{R}'^2_{ac} \nonumber 
\end{eqnarray}
where $\tilde{R}'^1_{ab}$ stands for $\tilde{R}' (\phi
_{1,a}-\phi_{1,b})$ etc..
The bilocal term thus has a scale invariant fixed form of order
$\epsilon^2$ and is  a well-defined function of
$\tilde{q}=q/\Lambda_l$ with no divergences.

More generally, we conjecture that there is a fixed asymptotic form
for all multilocal interactions $V^{(p)}$ which can be explicitly
written as a sum of properly rescaled multi-replica terms as
\begin{eqnarray}
V_l^{(p)}(\phi_1,..\phi_p,x_1,..x_p) 
= \Lambda_l^{D p} \sum_{c \geq 2}
\frac{\Lambda_l^{c(2-D)}}{c! T^c} \sum_{a_1,..a_c}
\tilde{V}_l^{(p,c)}(\{ \phi_{\alpha,a_i} \}_{i=1,..c}^{\alpha=1,..p} ,
\Lambda_l x_1,..\Lambda_l x_p)
\end{eqnarray}
where the number of replicas $c$ corresponds graphically to the number
of connected components. 
The consistency of the method demands that the $\tilde{V}_l$ flow to
well defined fixed points, perturbative in $\epsilon$. It is indeed
natural to conjecture that in this theory there is no wave function
renormalization.

We can now come back to the calculation of the correlation function.
Although for convenience we have computed ${\sf C}_{ab}^{q}$ from the theory 
with $l=\ln(q/\Lambda)$, this is unnecessary. As discussed above, the 
existence of a fixed point with well-defined functions 
of $\tilde{q}=q/\Lambda_l$ implies that the $\epsilon$ expansion
of ${\sf C}_{ab}^{q}$ at fixed $q \ll \Lambda$ should be of the form:
\begin{eqnarray}
{\sf C}_{ab}^{q}= \frac{1}{q^4} \Lambda_l^\epsilon \tilde{\Sigma}(\tilde{q})
\end{eqnarray}
where the dimensionless self energy $\tilde{\Sigma}(\tilde{q})
$ 
depends only on $\tilde{q}$ and $\epsilon$.
Since ${\sf C}_{ab}^{q}$ is $l$ independent, it implies, taking the
derivative, that $\tilde{\Sigma}(\tilde{q}) = C_\epsilon
\tilde{q}^{\epsilon}$ and thus
\begin{eqnarray}
{\sf C}_{ab}^{q} &=& C_\epsilon q^{-D} \label{unsurqd}\\
C_{\epsilon } &=& \frac{2 \pi^2}{9} \epsilon
\end{eqnarray}
to lowest order in $\epsilon $, as in 
\cite{giamarchi_vortex_long}. The form (\ref{unsurqd}) 
should be valid to all orders in $\epsilon$ if
the hypothesis about the fixed point formulated above are satisfied.

\subsection{FRG at finite temperature}
\label{finitetemp}

\subsubsection{Renormalization equations}

Since in (\ref{exact}) the terms in the exponentials containing the temperature
go to zero as $\Lambda_l^{D-2}$ one can first study the effect of temperature,
compared to $T=0$ by looking at the linear term. Up to 
order $T R$ and $R^2$ (i.e to one loop) the RG equation thus reads:

\begin{eqnarray}\label{FRGT}
\partial_{l} \tilde{R}_{l} (u)= \epsilon \tilde{R}_{l}(u) + \hat{T}_l
\tilde{R}''_l(u) +
\int_{0}^{l}dl'\,{\sf K}_{l'} \left( 
\frac{1}{2}\tilde{R}^{\prime \prime }_{l-l'}(u)^2 -
\tilde{R}_{l-l'}^{\prime \prime}(0) \tilde{R}^{\prime
\prime}_{l-l'}(u)\right) 
\end{eqnarray}
with 
\begin{eqnarray}
&& \hat{T}_l = - \partial G_l^{x=0} = 
- T \int_q \Lambda_l^{-2} c'( \frac{q^2}{2 \Lambda_l^2}) \\
&& = 2 T S_4 \Lambda_l^2 \int_{s>0} c(s) + {\cal O}(\epsilon)
\end{eqnarray}
In the case of sharp cutoff this equation has been studied in
\cite{balents,chauve_creep_long}.
It was found that at fixed $u$, $\tilde{R}_l(u)$ converges to
$\tilde{R}^*(u)$ but that 
temperature rounds the cusp of the $T=0$ solution in a boundary layer of
size $u \sim \hat{T}_l$. As in \cite{chauve_creep_long} 
we look for a solution of the form:
\begin{eqnarray}\label{anole}
\tilde{R}_{l} (0) - \tilde{R}_{l} (u) = -\tilde{R}_{l}'' (0)
\frac{u^{2}}{2} - {\sf K}
\frac{\hat{T}^{3}_{l}}{\epsilon ^{2} \chi^{2}} 
H_{l} (\frac{u\epsilon \chi }{\hat{T}_{l}})
\end{eqnarray}
Here, $\chi $ (of order $\epsilon ^{0}$) is defined 
by $\epsilon \chi =\tilde{R}^{*\prime \prime \prime } (0^{+})$ and thus
$H^{*\prime \prime \prime \prime }(0)=-1$. One has also 
$H_{l} (0)=H''_{l} (0)=0$.
Injecting (\ref{anole}) into (\ref{FRGT}) 
and identifying the leading order in
$\hat{T}_{l}$, one gets:
\begin{eqnarray}
\frac{x^{2}}{2}=H^{*\prime \prime}(x) + 
\frac{1}{2} \int_{l >0} K_{l } e^{4l }
H^{*\prime \prime}(xe^{-2l })^{2}
\end{eqnarray}
with $\int_{l >0} K_{l }=1$. This equation can be solved
iteratively in $n$ for $H^{*} (x)=\sum_{n>0}a_{n}x^{2n}/ (2n)!$. One has
$a_{1}=1$, $a_{2}=-3$ but higher $a_{n}$'s are non-universal. The
large $x$ behavior of $H^{*} $ is universal and given by $H^{*}  (x)\sim x
$. In the case of sharp cutoff one recovers~\cite{chauve_creep_long}
\begin{eqnarray}
H^{*\prime \prime } (x) = \sqrt{1+x^{2}}-1
\end{eqnarray}

This result should be further examined by consideration of consistency within
higher loop corrections, which goes beyond this paper.

The most important result of the section is that the following relation
between the finite temperature solution and the $T=0$ solution:
\begin{eqnarray}
\lim_{l \to +\infty} \hat{T}_{l} \tilde{R}''''_l(0) 
= C \tilde{R}^{* \prime \prime \prime}_l(0^+)^2
\label{important}
\end{eqnarray}
where $C=S_4$ holds irrespective of the cutoff function, and thus is determined
by the $T=0$ fixed point. This property will be used below.

\subsubsection{Calculation of universal susceptiblity fluctuations}

It was noted recently that a signature of glassy behaviour in a disordered
system was the large sample to sample fluctuations of the response to
external perturbations
\cite{mezard_parisi,hwa_fisher,nattermann_scheidl_revue}.
These are described by the following suceptibility:
\begin{eqnarray}
\label{defchi}
\chi = \frac{1}{T} \frac{1}{L^D} 
\int_{xy}
(\langle \partial_\alpha u^x \partial_\alpha u^y \rangle - \langle
\partial_\alpha u^x \rangle \langle \partial_\alpha u^y \rangle)
\end{eqnarray}
in a finite system of size $L$,
which measures the response in a given sample
to a field coupling to $\nabla u$ (e.g. the tilt or
compression a flux lattice, or the compression response). The 
$\langle X \rangle$ denotes the thermal averages in a given sample. These have 
been studied in connection with mesoscopic behaviour of disordered systems
\cite{emig_kardar}. Here we have considered only the trace of the response
tensor (extension being straightforward). 
To perform the calculation in the replicated theory we define:
\begin{eqnarray}
&& C^{\alpha \beta}_{ab} = \frac{1}{T} \frac{1}{L^D} \int_{xy}
\langle \partial_{\alpha} u_a^{x} \partial_{\beta} u_b^{y}  \rangle \\
&& C_{abcd} = \frac{1}{T^2} \frac{1}{L^{2D}} \int_{xyzt}
\langle \nabla u_a^{x}  \cdot \nabla u_b^{y} \nabla u_c^{z} 
\cdot \nabla  u_d^{t} \rangle 
\end{eqnarray}
We now compute respectively the first and second moment of the
sample to sample fluctuations of the susceptibility. They read:
\begin{eqnarray}
&& \overline{\chi} = C_{aa} -  C_{ab} = 1 \\
&& \overline{\chi^2} = C_{aabb} + C_{abcd} - 2 C_{aabc} 
\end{eqnarray}
where $\overline{X}$ denote disorder averages and $a,b,c,d$ take values
all distinct from each other. Note the well known
property \cite{brezin_sts} that the average susceptibility is
identical to the susceptibility of the pure system. We now compute
$C_{abcd}$ to lowest order in $\epsilon$. Only zero and one loop
graphs involving respectively $R_l''(0)$ and $R_l''''(0)$ contribute.
Interestingly, due to the quadratic nature of the term proportional to
$R_l''(0)$ the zero loop graphs cancel in $\overline{\chi^2}$, as can be
easily seen since $C_{abcd} = C_{ab}^2 + 2 \sum_{\alpha \beta} 
(C^{\alpha \beta}_{ab})^2$ for any 
gaussian theory (performing the Wick contractions). One is left
with:
\begin{eqnarray}
&& C_{abcd} =  ( n \delta_{abcd} - (\delta_{bcd} + \delta_{cda} +
\delta_{dab} +
\delta_{abc}) + (\delta_{ab} \delta_{cd} + \delta_{ac} \delta_{bd} +
\delta_{ad} \delta_{bc})) A R_l''''(0)
\end{eqnarray}
where $A = L^{-2D} \int_{w} (\int_{xy} \nabla G^{x-w} \cdot \nabla G^{y-w})^2$
and thus
\begin{eqnarray}
\overline{ (\Delta \chi )^2 } \equiv 
\overline{\chi^2} - \overline{\chi}^2 =  
A R_l''''(0) \sim R_l''''(0) C L^{4-D}
\end{eqnarray}
for a system of finite size $L$ (see also \cite{nattermann_scheidl_revue}
for a similar result in straight perturbation theory). 
Note that one can equivalently 
study the perturbation of an infinite system (i.e $L \to \infty$ first) 
by a periodic external field of wavevector $q_{ext}$.
In that case $A = q^{D-4}_{\rm ext}$. Thanks to the exact RG equations
at finite $T$ and substituting $l = \ln L$ we obtain the 
mesoscopic susceptibility fluctuations at low temperature as:
\begin{eqnarray}
\overline{ (\Delta \chi )^2 } =  C' \frac{L^{\theta}}{T}
\end{eqnarray}
where $\theta = D-2 + 2 \zeta$ is the energy fuctuation exponent
and $C' = {\cal O}(\epsilon^2)$ for a periodic system ($\zeta=0$)
and $C' = {\cal O}(\epsilon^{4/3}) \sigma^{2/3}$ for an interface
in random field disorder (see Section \ref{biased}).

This result, derived here through exact FRG calculation,
is consistent with the droplet picture \cite{fisher_huse}.
Indeed the second moment of
the susceptibility fluctuations is dominated by the rare configurations
of disorder (of probability $p_{\rm deg} \ll 1$)
with two almost degenerate (i.e within ${\cal O}(T)$ in energy)
ground states as follows:
\begin{eqnarray}
\overline{ (\Delta \chi )^2 } \sim p_{\rm deg} (\delta \chi_{\rm typ})^2
\end{eqnarray}
where $p_{\rm deg} \sim T/L^\theta$ and the typical fluctuation
is $\delta \chi_{\rm typ} \sim T^{-1} L^{-D} L^{2 D - 2 + 2 \zeta}$
from \ref{defchi}. One thus recovers the above result
since $\theta = D-2 + 2 \zeta$.

\subsection{Interface in a biased random field and toy model}
\label{biased}

In this Section we study the model (\ref{interface}) in the presence
of a mass term $m >0$, which confines the fluctuations of the displacement
$u^x$. We consider two cases (i) random field disorder (ii) periodic disorder. 
A physical realization of (i) consists in a domain wall separating the
$\pm$ phases in a ferromagnet, submitted to a random magnetic field.
The magnetic energy of the interface, assumed without overhangs, is:
\begin{eqnarray}
{\cal E}(u) = 2 \int d^D x \int_0^{u^x} du' h(u',x) 
\label{energy}
\end{eqnarray}
Thus the effect of the mass term corresponds 
to applying an additional field gradient $h(u,x) \to h(u,x) + m^2 u/2$.
Note that this field gradient can either stabilize ($m^2 >0$)
or destabilize ($m^2 <0$) the domain wall. We will study the
approach to the critical value $m^2 \to 0^+$. The case (ii) is
of interest when studying the competition between disorder and
e.g. a periodic potential. In the phase where the periodic potential
is relevant 
it is natural to approximate it by replacing it by an harmonic well 
(see e.g. \cite{mott_glass}).

Here we examine only ground state properties (zero temperature). 
We show that the
disorder induced fluctuations of the displacement $u^x$ is described, as
$m \to 0$ by a universal scaling function of the form:
\begin{eqnarray}
\overline{ u^q u^{-q} } = m^{- \alpha} F[c q^2/m^2]
\end{eqnarray}
which we determine to lowest order in $\epsilon=4-D$. Note that
$c$ can be measured from the thermal connected correlation, which
is unchanged by disorder for models like (\ref{interface}) 
which possess the statistical tilt
symmetry.

\subsubsection{RG equations in presence of a mass}

In this Section it is more convenient to use the RG equation resulting
from the multilocal expansion on $\hat{\cal V}$, which is local in $l$
to this order. Since we are studying $T=0$ we set $\hat{R}_l = R_l$ in
the following. The RG equation reads:
\begin{eqnarray}
&&  \partial_l R_l (u) = J_l \left( \frac{1}{2} R''_l (u)^2-R''_l (0)R''_l (u)\right)
\label{ldep} \\
&& J_l = - 2 \int_q \partial \overline{G}_l^q \overline{G}_l^q 
\end{eqnarray}

Using the action ${\cal S}_l$, the $T=0$ correlation function reads
to lowest order in $R$:
\begin{eqnarray}
\overline{ u^{q} u^{-q} }  = - R_l''(0)
\left( \frac{c(q^2/2\Lambda^2)}{c q^2+ m^2}\right)^2
\label{corrrfim}
\end{eqnarray}

It is easy to transform (\ref{ldep}) into the RG equation in the absence of 
a mass. Using the change of variable:

\begin{eqnarray}
R_l(u)=c^2 \Lambda_{t(l)}^{\epsilon-4\zeta} \tilde{R}_{t(l)}(u \Lambda_{t(l)}^{\zeta})
\end{eqnarray}
one finds that $\tilde{R}_l$ satisfies the $m=0$ flow equation:

\begin{eqnarray}
\partial_t \tilde{R}_t(x)=(\epsilon-4\zeta) \tilde{R}_t(x)+ \zeta x \tilde{R}'_t(x)
+ S_D \left( 
\frac{1}{2} \tilde{R}''_t (x)^2- \tilde{R}''_t (0) \tilde{R}''_t (x)\right)
\end{eqnarray}
provided that $t(l)$ satisfies
\begin{eqnarray}
\Lambda^{-\epsilon}_{t(l)} \frac{dt}{dl} = J_l \frac{c^2}{S_D}
\end{eqnarray}
which is integrated into:
\begin{eqnarray}
\frac{e^{\epsilon t_l}-1}{\epsilon}&=&\int_{0}^\infty q^{D-1}dq\, 
\frac{c(q^2/2)^{2}-c(q^2e^{2l}/2)^{2}}{(q^2+\alpha)^{2}}
\end{eqnarray}
with $\alpha=\frac{m^2}{c\Lambda^2}$. The function $t(l)$ is
increasing and bounded. Its limit $t(+\infty)=t_\infty$ for $D<4$
is given by:
\begin{equation}
\frac{e^{\epsilon t_\infty}-1}{\epsilon} = \int_{0}^\infty q^{D-1}dq\, 
 \frac{c(q^2/2)^{2}}{(q^2+\alpha)^{2}}
\end{equation}
and it diverges for $m \to 0$ as:
\begin{equation}
e^{\epsilon t_\infty} \sim  \epsilon \int_{0}^\infty q^{D-1}dq\, 
 \frac{1}{(q^2+\alpha)^{2}} \sim \alpha^{-\frac{\epsilon}{2}} (1-\frac{\epsilon}{2})
\frac{\frac{\epsilon \pi}{2}}{\sin \frac{\epsilon \pi}{2}}
\label{tinf}
\end{equation}
In $D=4$ it diverges as:
\begin{eqnarray}
t_\infty \sim \frac{1}{2} \ln(\frac{1}{\alpha})
\end{eqnarray}
We now distinguish the two cases.

\subsubsection{Random field}

In that case the correlations of the potential are 
$\overline{(W(r,u)-W(r',u'))^2}=-2 \delta ^D (r-r') R(u-u')$
with (from (\ref{energy}), $R(u) \sim - \sigma |u|$ at large 
$u$. In the massless case it is known that the FRG to one loop
reproduces the purely dimensional result $u_{0}u_{r}\sim \sigma^{2/3} c^{-4/3} 
r^{2(4-D)/3}$ with a roughness exponent $\zeta= (4-D)/3$. From this
we expect, in the massive case, the small $m$ behaviour:
$$
u \sim \sigma^{1/3} m^{-\epsilon/3} c^{-D/6}
$$
It is known \cite{fisher_functional_rg} 
that one must fix $\zeta=\epsilon/3$ to obtain a reasonable
fixed point. From the above equation, the (reduced) correlator of the force 
$\Delta_t (x)=- \tilde{R}''_t(x) S_D$ then satisfies 
$$
\partial_t  \Delta_t (x)=\frac{\epsilon}{3}\left( x \Delta_t 
(x)\right)'-\frac{1}{2}{\left( \Delta_t (x)-\Delta_t (0)\right)^2}''
$$
and flows for $t \rightarrow \infty$, to a fixed point 
$\Delta_\infty (x) $ given in terms of a function $y(x)$
\cite{fisher_functional_rg} implicitly
defined as \cite{chauve_creep_long}: 
$$
\left \{
\begin{array}{ccc}
\Delta_\infty (x)&=&\Delta_\infty (0) 
y(x\sqrt{\frac{\epsilon}{3\Delta_\infty (0)}})\\
\Delta_\infty (0)&=&\left( 
\frac{\epsilon}{24 \gamma^2}\right)^{1/3}
\left(\int_{-\infty}^{+\infty} dx \, \Delta_0(x) \right)^{2/3}\\
\frac{x^2}{2}&=&y(x)-1-\ln y(x)
\end{array}\right.
$$
with $\gamma = \int_0^1 dy\, \sqrt{y-1-\ln y} = 0.5482228893..$
Note that $\int_{-\infty}^{+\infty} dx \, \Delta_t(x)$ is $t$
independent and thus equal to 
$\int_{-\infty}^{+\infty} dx \, \Delta_0(x) = - \frac{S_D}{c^2} \int R''_0
= 2 \sigma \frac{S_D}{c^2}$.

Putting this together with (\ref{corrrfim}), (\ref{tinf}) yields the
result:
\begin{eqnarray}
&& \overline{ u^{q} u^{-q} }  = \sigma^{2/3} m^{-D - 2 \zeta} c^{D/6}
F_D[c q^2/m^2] \nonumber \\
&& F_D[x] = C_D \frac{1}{(1+x)^2} + \mbox{h.o.t} \label{rfim} \\
&&  C_D = \left( \frac{(4\pi)^{D/2}}{6 \Gamma(\frac{\epsilon}{2})
\gamma^2} \right)^{1/3} \nonumber
\end{eqnarray}
with $\zeta=\epsilon/3$. Note that the universal scaling function must behave
at large $x$ as $F_D[x] \sim x^{\zeta - D/2}$, and is determined here only
to order $0$ in $\epsilon$.

From this one also finds the local fluctuation:
\begin{eqnarray}
&& \overline{ (u^x)^2}  = \epsilon^{-2/3} 6^{-1/3} 
\left( \frac{\epsilon \Gamma(\frac{\epsilon}{2}) }{
(4\pi)^{D/2} \gamma} \right)^{2/3}
\sigma^{2/3} m^{-2 \zeta} c^{-D/3}
\label{rfim2}
\end{eqnarray}
which is also universal. The fact that this quantity is dominated by
large scale fluctuations can be seen from the convergence of
the integral ($\zeta >0$).

The calculation can also be performed exactly in $D=4$. One finds:
\begin{eqnarray}
&& \overline{u^{q}u^{-q}} = \sigma ^{2/3} c^{2/3}
m^{-4} \left( \ln \frac{1}{m } \right)^{-1/3}
\left(\frac{4 \pi ^{2}}{3 
\gamma^{2}} \right)^{1/3}
\frac{1}{(1+\frac{cq^{2}}{m^{2}})^{2}}
\end{eqnarray}

The values we find for $C_D$, which vanishes as $C_D \sim 3.5246\, 
\epsilon^{2/3}$ 
as $D \to 4^{-}$ are as follows
$C_3 \approx 2.40653$, $C_2 \approx 1.91006$, $C_1 \approx 1.30416$,
$C_0 \approx 0.82157$, remarkably close to 
the exact result in $D=0$ \cite{ledou_monthus} $C_0 = 1.05423856519$.

It is also useful to compare these results with the Gaussian variational method
with replica symmetry breaking. Extending the calculation of
\cite{mezard_parisi}
to the non zero mass case, which is done in Appendix~\ref{variational}, 
we find the
same form as (\ref{rfim}) with
\begin{eqnarray}
&& F_D[x] = C'_D [ \frac{1}{(1+x)^2} + 
\frac{\theta(2 - \theta)}{4} 
\frac{1}{1 + x} \int_1^{+\infty} 
\frac{dy}{y^{1+\frac{\theta}{2}}} \frac{y-1}{y+x} ] \\
&&  C'_D/C_D = (\frac{12}{\pi} \gamma^2)^{1/3} = 1.04708
\label{rfimmp}
\end{eqnarray}
and $\theta=(6 -\epsilon)/3$ and $\zeta=\epsilon/3$. 
Since as $\epsilon \to 0$ for fixed $x$ the second integral is subdominant, the
leading order in $\epsilon$ are identical and the amplitude of the
RSB solution compared to the FRG solution is $C'_D/C_D$
as $\epsilon \to 0$.

\subsubsection{Periodic case}

In the case of a periodic system with period $a$, one gets a fixed
point function with $\zeta=0$ which reads:
\begin{equation}
\Delta_\infty (u)=\frac{\epsilon}{6}\left( \frac{a^{2}}{6}-u (a-u)\right)
\end{equation}
It yields:
\begin{eqnarray}
\overline{u^{q}u^{-q}} &=& a^{2}c^{D/2}m^{-D}
\frac{(4\pi)^{D/2}}{36\Gamma [\frac{\epsilon }{2}]} 
\frac{1}{(1+\frac{c q^{2}}{m^{2}})^{D/2}}\\
\overline{ u_{r}^2 }  &\sim& \frac{\epsilon}{36} a^{2} \ln(\frac{1}{m})
\end{eqnarray}
and in $D=4$:
\begin{eqnarray}
&& \overline{u^{q}u^{-q}} = \frac{2 \pi^2 }{9} a^{2}c^{2}
\frac{m^{-4}}{\ln \frac{1}{m }}
\frac{1}{(1+\frac{cq^{2}}{m^{2}})^{2}}
\end{eqnarray}

\section{towards two loop FRG}

\label{2l}

The exact RG method allows to compute quantities
beyond the lowest order in $\epsilon$. It can be carried either at $T \to 0$
for fixed system size ($T=0$ limit) or at finite $T$.
Solving the exact RG equation at $T=0$ requires to follow
non analytic functions. This is a difficult question, e.g. distinguishing the
various cumulants in the local part demands a special procedure that
we have developed. This is discussed in \cite{us_part2,kay_dynamics}.

These problems do not arise at $T>0$ where the singularity is smoothed
within a boundary layer (at one loop see Section \ref{finitetemp}).
In the Appendix \ref{totwoloops} we have used the exact RG flow 
to third order (given in \ref{rgu}) and obtain the two loop
exact FRG equation for the second cumulant $R_l(u)$ at $T>0$.
At large $l$ the effective temperature $\hat{T}_l \to 0$ and
one recovers an ``effective'' zero temperature equation. This equation 
differs from the one obtained in \cite{larkin_2loops} as it contains
a new ``anomalous'' term of the form $\lambda R'''(0^+)^2 R''(u)$. We find that the
coefficient of this term is universal with $\lambda=1/2$. Interestingly this
value is consistent with the renormalizability arguments given in \cite{kay_dynamics}.
The question of calculation of correlations and of their 
universality to two loops is rather delicate 
\cite{us_part2}. Let us mention here that our exact RG
fixed point equation depends only on one non universal coefficient
$\overline{K}^C$, which vanishes in the case of sharp cutoff. The solutions of
this equation, given in Appendix \ref{totwoloops}, exhibit the property of
a non zero value of $R'(0^+)$, referred to as a supercusp
since it is a stronger non analyticity than the one loop one ($R'''(0^+) \neq 0$).
This feature is unpleasant as it naively yields (by perturbative expansion)
additional divergences as $T^{-1/2}$, a sign of possible 
fractional dependence in $\epsilon$
(a related phenomenon is discussed in \cite{balents_fisher}). 
On the other hand since its magnitude is proportional to $\overline{K}^C$
it is absent in sharp cutoff calculations.
This problem and its relation to the structure of the boundary layers
at high orders is further examined in \cite{us_part2}.

\section{Conclusion}
\label{conclusion}

In this paper we have introduced a systematic method which
turns the exact, though abstract, RG functional equation of Wilson-Polchinski
into a tool for concrete perturbative calculations to any number of loops
using arbitrary cutoff functions. The strategy was to explicitly integrate
out all non local interactions, which can be expressed in terms of the
local part alone,
order by order in the local part. In the process we have preserved
the exactness and the controlled nature of the original
Wilson-Polchinski 
equation. Indeed, no approximation was made, and the resulting RG equation for
the local part, as well as the expressions for the nonlocal ones
and for the correlation functions, are formally exact order by order in an
expansion in the local part. This expansion will be useful for theories
where the local part is small, i.e when it is controlled by a small 
parameter (e.g. the shift from the upper critical dimension) and
when the RG equation admits a perturbative fixed point in
this parameter. We have considered here theories with a bare local interaction
and a fixed point for the local part, e.g. as in the $O(n)$ model, but the
method is more general and can be extended to theories where the bilocal part 
of the interaction serves as the small parameter, e.g. 
for self avoiding manifolds
\cite{wiese}. In a sense, the
exact RG in the operational form presented here directly translates the ideas 
of Wilson and provides explicit checks of universality.

In addition to presenting the method formally to all orders, we have 
derived the {\it explicit} RG equation for the local part up to third order.
Further expanding in the number of loops, we have explicitly given the
coefficients 
up to two loops and third order. Two distinct, although equivalent, methods
have been presented, depending on whether one considers the Wick ordered
functional or not. Each method has its advantages: 
the Wick ordered method yields apparently simpler
(less nonlocal) RG equations, but it is not always the most adequate
(e.g. for the finite $T$ one loop FRG analysis).
Although the present paper contains all the material necessary for
two loop applications (e.g. for $O(n)$ and for the FRG) we have preferred
to defer giving the detailed calculations and results to a companion
paper~\cite{us_part2}. In particular we have sketched here the simple
extensions needed to deal with the so-called wave function renormalization
which arises in e.g. the $O(n)$ model to two loops, examined in more details
in~\cite{us_part2}.

We have thus considered here mostly one loop applications. The first one was a 
simple check to recover the one loop exponents of the $O(n)$ model.
The second application was to the theory which describes elastic systems
in random potentials. It was previously analyzed through simpler
Wilson momentum shell integration~\cite{fisher_functional_rg} 
but the rather unusual nature
of the theory ($n \to 0$ limit, non-analyticity) made it important to
verify explicitly 
that the results are universal. Also universality in disordered systems
is rather less established than in pure systems, especially in the $T=0$ limit
where it is known to fail in some cases. Thus we first derived the $T=0$ 
one loop RG equation for the second cumulant function $R(u)$ in
arbitrary cutoff scheme and 
found that its coefficients are universal to this order. This yields the
universality to ${\cal O}(\epsilon)$ of the roughness exponent 
$\zeta$ of pinned
interfaces. In the periodic case, we also explicitly verified that the
correlation function contains a universal amplitude. Similarly, we 
computed the scaling function of the ground state deformations
of a confined interface in a random field and found a universal result.
This quantity can be experimentally measured in disordered magnets in
the presence of a small additional field gradient.

Although temperature is formally irrelevant (the dimensionless
temperature flows 
to zero) it is well known to be ``dangerously'' so. Our exact FRG at $T>0$
shows that although the ``boundary layer'', i.e the detailed asymptotic 
form of the cumulant $R(u)$ for $u \sim T_l $ is nonuniversal, some of
its features are universal, and in particular we were able to extract from
it the universal divergence of the mesoscopic fluctuations of the 
suceptibility $\Delta \chi$. The divergence of this quantity, which is
dominated by rare almost degenerate low energy configurations,
is an accepted unambiguous measure of ``glassiness'' in a disordered system
and is measured in experiments, e.g. in microsize vortex systems.

Some of the peculiar features of the theory of pinned elastic systems 
have been also discussed. We have found it useful to give a detailed
diagrammatic 
proof of the triviality of naive perturbation theory, as we have not
seen it explicitly 
in the litterature (though more general statements about dimensional reduction
appear in a number of other works). We have discussed how the non analytic
nature of the theory yields non trivial results.

Pinning of disordered media thus provided us here with one example of
a problem 
where exact renormalization is needed to get insight, as no
field theoretical description is yet available. The reason for it is
that one must follow in principle a rather complicated object, the
full probability distributions or the disorder, or equivalently the whole
series of cumulants. The method seems thus promising for other problems
with similar features, such as random Sine Gordon models~\cite{carpentier}.
It is interesting to note that while presenting the FRG method, Fisher
pointed out (Ref. 12 in \cite{fisher_functional_rg}) that the
momemtum-shell RG ``suffers from pathologies due to the sharp
cutoff'', and that the cusp in $R (u)$ ``requires a careful analysis
of the full renormalization group''. Through the use of the exact RG
method presented here, we provide a simple way 
to integrate explicitely and exactly what is left aside
in the traditional RG. We are thus able to control the
approximations of former approaches and to perform new
calculations. Furthermore, thanks to our general framework expressed
in terms of {\it any} cutoff function, the universality of the results
is checked.

Let us close by noting that the application of the multilocal solution
to the exact RG equation seems promising also to study other disordered 
problems, or even give a new perspective on simpler pure problems.
For instance one could apply it to wetting problems taking into account
the nonlinear part, or to the roughening problems to improve on
previous analysis using uncontrolled projections methods \cite{anusha-jpb}
The multilocal expansion allows also
interesting extensions to theories with bilocal bare action, such as 
polymers, mutually interacting or with disorder. 
Finally, it is also worth studying more closely the set of exact solutions 
to the Polchinski equation presented in this paper (Appendix \ref{examples}).
Some of these extensions will be explored in future publication.

\appendix

\section{Invariance properties of generating functional and
renormalization equation}
\label{invariance}

In this Appendix we give a concise derivation of the
exact invariance properties of the generating functional of correlation
functions under coarse graining. These properties provide 
the basis for developing exact renormalisation procedures
of the Polchinski type. In the second part we generalize the framework
to include additional field transformations, such as rescaling. This
extended framework is suitable for theories where
wave function renormalization must be included
(see \cite{us_part2}). 

\subsection{Invariance under coarse graining}
\label{coarsegrain}

We use only the two following properties of Gaussian averages.
The notations are the same as in the body of the paper.
First, transformation under a change of variable 
$\phi \rightarrow \phi +\psi $ for any field $\psi $ 
in the functional integration over $\phi $ yields:
\begin{eqnarray}\label{gaussshift}
\left[{\cal A} (\phi ) \right]_{G}=e^{-\frac{1}{2}\psi :G^{-1}:\psi }
\left[e^{-\psi :G^{-1}:\phi } {\cal A} (\phi +\psi ) \right]_{G}
\end{eqnarray}
for any functional ${\cal A} (\phi )$. We will also use the
composition property:
\begin{eqnarray}
\left[\left[{\cal A}(\phi_1+\phi_2
) \right]_{G_{2}} \right]_{G_{1}} = \left[{\cal A}(\phi) 
\right]_{G_{1}+G_{2}}
\end{eqnarray}
where in the l.h.s. the average over $\phi _{i}$ is performed using Gaussian
correlations $G_{i}$.

Using successively the shifts
$\phi_{1}\rightarrow \phi_{1}-G_{2}:J$ and
$\phi_{2}\rightarrow \phi_{2}+G_{2}:J$ yields the fundamental relation 
\begin{eqnarray}
\left[ e^{J:\phi -{\cal V}(\phi )}\right]_{G_0}&=&
\left[\left[ e^{J:(\phi_{1}+\phi _{2}) -{\cal V}(\phi_{1}+\phi _{2}
)}\right]_{G_{2}}\right]_{G_{1}}\\
&=&e^{-\frac{1}{2}J:(G_{2}+G_{2}:G_{1}^{-1}:G_{2}):J}
\left[ e^{J:(1+G_{2}:G_{1}^{-1}):\phi_{1}}
\left[ e^{-{\cal V}(\phi_1+\phi _{2})}\right] _{G_2}\right] _{G_1}
\end{eqnarray}
where we denoted $G_{2}=G_0-G_{1}$. Thus, if ${\cal V}_{1}$ is
coarse--grained transformed
of the interaction ${\cal V}$, defined by
\begin{eqnarray}\label{vv1}
e^{-{\cal V}_{1}(\phi_1)}=\left[ e^{-{\cal V}(\phi_1+\phi
_{2})}\right] _{G_0-G_{1}}
\end{eqnarray}
then one has
\begin{eqnarray}  \label{fund2}
\left[ e^{J:\phi -{\cal V}(\phi )}\right]_{G}&=&
e^{\frac{1}{2}J:(G_0-G_0:G_{1}^{-1}:G_0):J}
\left[ e^{J:G_0:G_{1}^{-1}:\phi_{1}}
e^{-{\cal V}_{1}(\phi_1+\phi _{2})}\right] _{G_1}
\end{eqnarray}

We now use this property of Gaussian integrals as follows. One defines
a family of actions ${\cal S}_{G} (\phi )$ and
their associated generating functional $W_{G}(J)$
\begin{eqnarray}
{\cal S}_{G} (\phi )=\frac{1}{2}\phi :G^{-1}:\phi +{\cal V}_{G} (\phi
)\qquad W_{G}(J)=\ln \left[e^{J:\phi -{\cal V}_{G} (\phi)} \right]_{G}
\end{eqnarray}
They are indexed by the matrix $G$ and we choose them to be related 
by the coarse graining operation (\ref{vv1}) where $G$ plays the role
of $G_1$, namely:
\begin{eqnarray}\label{vnew}
e^{-{\cal V}_{G}(\phi)}=\left[ e^{-{\cal V}_{G_0}(\phi+\psi
_{2})}\right] _{G_0-G}
\end{eqnarray}
or, equivalently in a differential form, the ${\cal V}_{G}$ satisfy
the ``RG equation'':
\begin{eqnarray}\label{coarsegraining}
\frac{\delta }{\delta G}e^{-{\cal V}_{G}(\phi)}=\frac{1}{2}
\frac{\delta ^{2}}{\delta \phi \delta \phi } e^{-{\cal V}_{G}(\phi)}
\end{eqnarray}
obtained by differentiating (\ref{vnew}) with respect to $G$.
The coarse--graining equation (\ref{coarsegraining}), 
read along a given path $l\mapsto G_{l}$, is the
Polchinski equation in its ``diffusive'' form (\ref{diffusion}).

It is easy to see from (\ref{fund2}) that this choice of
a family ${\cal V}_{G}$ implies the property:
\begin{eqnarray}
\tilde{W}_{G}(J:G':G^{ -1})\mbox{ independent of }G
\end{eqnarray}
where we have defined the interaction part
$\tilde{W}_{G} (J)=W_{G}(J)-\frac{1}{2}J:G:J$. It
allows to relate correlations within any member of the
family ${\cal S}_{G}$, i.e under coarse graining.

\subsection{Generalization including rescaling and change in
Gaussian part}\label{zphi}

The previous properties can be extended to a larger set of
transformations which include simultaneous (i) coarse graining 
(ii) linear transformation of the fields (iii) redefinition
of the Gaussian part (such as needed to absorb its possible 
renormalization). It is based on the following properties 
of Gaussian integrals. One defines:
\begin{eqnarray}
W_{G,{\cal V}}(J) = \ln [ e^{J : \phi - {\cal V}(\phi)} ]_{G}
\end{eqnarray}
The first property correspond to performing an arbitrary
linear transformation on the field:
\begin{eqnarray}
W_{G,{\cal V}}(J) = W_{M^{-1} :G :M^{-1},{\cal V}(M :\phi)}(J: M)
\end{eqnarray}
valid for any $G,{\cal V},J,M$.
The second property is simply the identity obtained when 
redistributing the Gaussian part:
\begin{eqnarray}
W_{G,{\cal V}}(J) = - \frac{1}{2} {\rm Tr} \ln (1+ H: G) +
W_{(G^{-1} + H)^{-1},{\cal V}(\phi) - \frac{1}{2} \phi : H :\phi}(J)
\end{eqnarray}
valid for any $G,{\cal V},J,H$.

First, let ${\cal V}_G(\phi)$ satisfy the RG equation (\ref{coarsegraining}).
Then from the previous Section we know that 
$W_{G,{\cal V}_G}(J: G': G^{-1}) - \frac{1}{2} J: G': G^{-1}: G': J$ is
independent of $G$ for any $J, G'$.
Setting $G=G_l$ and $G'=G_0$ one gets the Polchinski equation and
one can compute $W(J)$. One now defines
\begin{eqnarray}
{\cal V}_{G,M}(\phi) = {\cal V}_{G}(M :\phi) + \frac{1}{2} 
\phi: (M: G^{-1}: M - G^{-1}) :\phi
\end{eqnarray}
Using the above properties one has
\begin{eqnarray}\label{invwgvgm}
W_{G,{\cal V}_{G,M}}(J :G': G^{-1}: M) - \frac{1}{2} J :G': G^{-1}: G' :J 
- \frac{1}{2} {\rm Tr} \ln (M^{-1} :G^{-1} :M^{-1}: G) 
\end{eqnarray}
is independent of $G$ and $M$, for any $G',J$.

We have used the two invariances choosing 
$H = M^{-1} :G^{-1}: M^{-1} - G^{-1}$
leading to the intermediate formula:
\begin{eqnarray}
W_{G,{\cal V}}(J) = - \frac{1}{2} {\rm Tr} 
\ln (M^{-1} :  G^{-1} :M^{-1} :G) +
W_{G, {\cal V}(M :\phi) + \frac{1}{2} \phi :(M: G^{-1} :M - G^{-1}):
\phi}(J: M)
\end{eqnarray}

Defining a new family of functional indexed by $l$ as:
\begin{eqnarray}
{\cal V}_{l}={\cal V}_{G_l,M_l}
\end{eqnarray}
and symmetric matrices $G,M$ one finds that the functional 
${\cal V}_{l}$ now satisfies a new RG equation
\begin{eqnarray}
&& \partial_l {\cal V} 
= \frac{\partial {\cal V}}{\partial \phi} :M^{-1} :\partial M :\phi
+ \phi :G^{-1} :M^{-1} :\partial M :\phi \\
&& -
\frac{1}{2} {\rm Tr}(
\partial G : \frac{\partial^2 {\cal V}}{\partial \phi \partial \phi} )
+ \frac{1}{2} \frac{\partial {\cal V}}{\partial \phi} : \partial G :
\frac{\partial {\cal V}}{\partial \phi}
\end{eqnarray}
which contains additional terms. In particular the quadratic piece
can be used to absorb ``wave function renormalization'' terms,
so as to keep ${\cal V}_{l}$ small. Once this equation is solved
the correlations can be related within any of the corresponding
${\cal S}_l$ theories using the above invariance property (\ref{invwgvgm}) of
$W_{G,{\cal V}_{G,M}}(J)$. This will be further exploited in
\cite{us_part2}.

\section{General properties and exact solutions of Polchinski equation}
\label{examples}

Let us first mention a few general properties of (\ref{poleq},\ref{diffusion}).
For the class of cutoff functions (\ref{examplecutoff}) used in
practice, the diffusion tensor in
(\ref{diffusion}) is positive $c' (s)\leq 0$ (but not definite
positive since there exists modes with $\partial_{l}G_{l}^{q}=0$).
There are some exactly formally conserved quantities, such as
$\int_{\phi }
e^{-{\cal V}_{l} (\phi )}$ and $\left[e^{-{\cal V}_{l} (\phi )}
\right]_{G_{l}}$.
Since (\ref{diffusion}) is a diffusion equation, it satisfies a 
${\rm H}$-theorem of increase of the ``entropy'' 
${\sf S}_{l} = \int_{\phi } {\cal V}_{l}
(\phi )e^{-{\cal V}_{l} (\phi )}$, which flows as 
$\partial_{l} {\sf S}_{l} = -\frac{1}{2} \int_{\phi } \frac{\delta
}{\delta \phi }e^{-{\cal 
V}_{l} (\phi )} :\partial_{l} G_{l}:\frac{\delta }{\delta \phi }e^{-{\cal
V}_{l} (\phi )} \geq 0$ and is compatible with the fact that RG
trajectories do not have limit cycles. 
Finally, since 
(\ref{diffusion}) is a linear equation, if we now a set of
solutions ${\cal V}^{\alpha }_{l} (\phi )$, then any superposition
such as 
\begin{eqnarray}
{\cal V}_{l} (\phi )=-
\ln \sum_{\alpha }c^{\alpha }e^{-{\cal V}_{l}^{\alpha } (\phi )}
\end{eqnarray}
is also solution.

This can now be used to construct non trivial exact solutions
to the Polchinski equation. The simplest family of exact solutions is of
course the quadratic potential, for which one finds the solutions
\begin{eqnarray}\label{polm}
{\cal V}_{l} (\phi )=\frac{1}{2} (\phi-\psi )
:M_{l}:(\phi-\psi ) - \frac{1}{2}{\rm Tr
}\ln M_{l}\\
M_{l}=\left( M_{0}^{-1}+G_{0}-G_{l} \right)^{-1}
\end{eqnarray}
where $\psi $ is an $l$-independent
field, with $\langle \phi \rangle _{{\cal S}}=
(1+M^{-1}:G^{-1})^{-1}:\psi$.

A much less trivial family of {\it exact solutions} of Polchinski
equation is obtained by superposition
of gaussians, i.e of quadratic potentials. It reads:
\begin{eqnarray}
{\cal V}_{l} (\phi )=-\ln \,\,\sum_{\alpha }c^{\alpha }\,\,\,e^{
-\frac{1}{2}(\phi-\psi^{\alpha } ):M^{\alpha }_{l}:(\phi-\psi^{\alpha
} ) +\frac{1}{2}{\rm Tr}\ln M^{\alpha }_{l}} 
\end{eqnarray}
with arbitrary constant coefficients $c^{\alpha }$ and each $M^{\alpha }_{l}$
satisfies (\ref{polm}). This is somewhat 
reminiscent of a decomposition into ``pure states'' and is clearly of interest
to describe low temperature states in pure models (in phases with broken symmetry)
or in disordered models and glasses (with many metastable states). It is
an interesting question to ask, quite generally, whether this family of solutions
can in some cases be an {\it attractive}
manifold in a larger functional space, or whether one can carry perturbation
around this subspace. These and related issues will be discussed in a 
future publication \cite{us_inprep}

A generic property of these solutions to Polchinski equation
is to generate cusp singularities separating the ``pure states''.
This can be seen directly above since negative curvatures tend to increase
in absolute value (see (\ref{polm}) and is presumably a very general
mechanism. It can also be seen on the simple example of the
zero-dimensional toy model. There, the field $\phi $ is a
real number and
\begin{equation}
Z = \int_{-\infty }^{\infty } d\phi \, e^{ -\frac{\phi ^{2}}{2G}-V (\phi )} 
\end{equation}
where $V$ is an arbitrary function. One can introduce $G_{l}=G-l$, 
and $V_{l} (\phi )$
for this model, verifying
\begin{equation}
\partial_{l}V_{l} (\phi )=\frac{1}{2}\left(V_{l}'' (\phi ) -V'_{l} (\phi
)^{2} \right)
\end{equation}
with initial condition $V_{0} (\phi )=V (\phi )$, and one can
integrate up to $l=G$. One has $e^{-V_{l} (\phi )}=\left[e^{-V (\phi
+\psi )} \right]_{l}$. 
The evolution of $V_{l} (\phi )$ is that the curvature $M_{l}=V''_{l}
(0)$, which would obey (\ref{polm}) $M_{l}=M_{0}/ (1+M_{0}l)$ for a
quadratic hill or well, 
diverges at a finite $l$ for maxima and decreases as $1/l$ for
minima of $V (\phi )$. Thus the landscape $V_{l} (\phi )$ develops
cusps, encoding for discontinuities in the force $-V' (\phi )$.

In the case of a periodic landscape, the natural
superposition of gaussian solutions is the {\it Villain potential}
$V (\phi )=-\ln \sum_n c e^{- (\phi -n)^{2}/ (2l)}$. In these
Sine gordon type potential, as well as in the $2D$ XY model, it is
a well known property that the renormalized potential converges towards
the Villain form at low temperature as found in \cite{jkkn} from the
Migdal Kadanoff RG (see also more recently \cite{anusha-jpb}).
The detailed behaviour of the RG flow can be studied in a more controlled
way using the method presented of this paper \cite{us_inprep}

\section{Multilocal expansion and higher order RG equation}
\label{higher}

In this Appendix we derive the systematic multilocal expansion
and obtain the RG equation to higher orders. We give a detailed
presentation for the functional $\hat{\cal V}_{l}(\phi)$, which
is simpler, and give explicitly the corresponding RG equation 
to order $\hat{\cal V}_{l}^3$ and up to two loops. Then we 
simply sketch the result for the same procedure applied
to the functional ${\cal V}_{l}(\phi)$, which is more
involved and will be presented in \cite{us_part2}.

\subsection{Multilocal expansion for $\hat{\cal V}$}

\label{vhatmulti}

The tadpole-free functional $\hat{\cal V}_{l}(\phi)$, defined in (\ref{hatv}),
can be written as a sum of multilocal interactions
\begin{eqnarray}
\hat{\cal V}_{l}(\phi)=\sum_{p>0}\int_{x_{1}..x_{p}}
V^{(p)}(\phi_{x_{1}}.. \phi_{x_{p}},x_{1}.. x_{p})
\label{decomposition}
\end{eqnarray}
Note that we are not even assuming here translational invariance.
The translationally invariant case discussed in (\ref{multiloc}) 
can be recovered 
by setting $V^{(1)}_l(\phi_1,x_{1})=\hat{U}_l(\phi_1)$,
$V^{(2)}_l(\phi_1,\phi_2,x_{1},x_{2})=\hat{V}_l(\phi_1,\phi_2,x_{1}-x_{2})$
etc.. Since we want to impose that each $V^{(p)}$, $p>1$, has zero local part
(this is sufficient for our purpose), we define (extending
(\ref{projo1},\ref{projo2})) 
respectively the projection operator $\overline{P_1}$ which 
projects a $p$-local interaction on a local one, and the
projection operator $P_1$ which transform a $p$-local interaction
into another $p$-local interaction as:
\begin{eqnarray}
&& (\overline{P}_1 A)(\phi,t) = \int_{x_{1}.. x_{p}} 
\delta (t-\frac{x_{1}+\dots +x_{p}}{p}) A(\phi,..\phi,x_1,..x_p) \\
&& (P_1 A)(\phi_1,..\phi_p,x_1,..x_p)
= \delta (x_{1}-x_{2})\dots \delta (x_{1}-x_{p})
\\ && \times \int_{y_{1}\dots y_{p}}
\delta (x_{1}-\frac{y_{1}+\dots +y_{p}}{p})
A(\phi _1\dots \phi _p, y_{1}\dots y_{p}) \nonumber
\end{eqnarray}
The property 
\begin{eqnarray}
\int_{x_1,..x_p} (P_1 A)(\phi_{x_1},..\phi_{x_p},x_1,..x_p)=
\int_t  (\overline{P}_1 A)(\phi_t,t) =
\int_{t} A(\phi_{t},..\phi_{t},t,..t)
\end{eqnarray}
ensures that one can choose the $V^{(p)}$, $p>1$ in the decomposition 
(\ref{decomposition}) to have no local part, i.e:
\begin{eqnarray}
P_1 V^{(p)}_l = 0 \qquad \overline{P}_1 V^{(p)}_l = 0
\end{eqnarray}
for any $l$ by applying $P_1$ and $1-P_1$ act on both sides of
the Polchinski equation.

Since the modified Polchinski equation (\ref{polhat}) concatenates two
operators, 
it is then easy to see that if the $V^{(p)}_l$ satisfy the
following set of equations:
\begin{eqnarray}
&& \partial_{l} V^{(1)}(\phi,t) = \frac{1}{2}
\sum_{p>0}\sum_{q=1}^{p-1} \int_{x_{1}\ldots x_{p}}
\delta (t-\frac{x_{1}+\dots +x_{p}}{p})
e^{\partial^{1\dots q}G_{l}\partial^{q+1\dots p}}
\partial^{1\dots q}\partial G_{l}\partial^{q+1\dots p} \nonumber \\
&&
V^{(q)}
(\phi_{x_{1}}\ldots \phi_{x_{q}},x_{1}\ldots x_{q})
V^{(p-q)}
(\phi_{x_{q+1}}\ldots \phi_{x_{p}},x_{q+1}\ldots x_{p})|_{\phi_{i}=\phi}\\
&& \partial_{l} V^{(p)}(\phi_{x_{1}}\ldots \phi_{x_{q}},x_{1}\ldots x_{q})
= \frac{1}{2} {\bf S} (1-P_{1}) 
\sum_{q=1}^{p-1} 
e^{\partial^{1\dots q}G_{l}\partial^{q+1\dots p}}
\partial^{1\dots q}\partial G_{l}\partial^{q+1\dots p} \nonumber \\
&&
V^{(q)}
(\phi_{x_{1}}\ldots \phi_{x_{q}},x_{1}\ldots x_{q})
V^{(p-q)}
(\phi_{x_{q+1}}\ldots \phi_{x_{p}},x_{q+1}\ldots x_{p})\mbox{ for }p>1
\end{eqnarray}
then (\ref{polhat}) is obeyed by $\hat{\cal V}_{l}(\phi)$. Since we
prefer to work with symmetric functions we have defined the symmetrization
operator:
\begin{eqnarray}
{\bf S}B^{(p)}(\phi _1\dots \phi _p, x_{1}\dots
x_{p})=\frac{1}{p!}\sum_{\sigma \in \Sigma _{p}}B^{(p)}(\phi _{\sigma
(1)}\dots \phi _{\sigma (p)}, x_{\sigma (1)}\dots x_{\sigma (1)})
\end{eqnarray} 
we have also defined the following shorthand notations:
\begin{eqnarray}
\partial^{1\dots q}\partial G
\partial^{q+1\dots p}=\sum_{\alpha=1}^{q}\sum_{\beta=q+1}^{p}\partial
G^{x_{\alpha}x_{\beta}}_{i j }\partial_{i}^{\alpha}\partial_{j }^{\beta}
\end{eqnarray}

It is easy to see that if $V^{(1)}$ is considered formally
as ``small'' in some sense (e.g. controlled by a small parameter such
as $\epsilon$) then one can integrate exactly these equations order 
by order in $V^{(1)}$ and check that 
$V^{(p)} = {\cal O}({V^{(1)}}^p)$. More precisely, to
a given order one can exactly integrate the equations for higher point
functions and reduce to a single equation for $V^{(1)}$.
This is the procedure that we now follow. The
structure to the lowest order $O({V^{(1)}}^2)$ is simply a closed equation
for $V^{(1)}$ of the schematic form:
\begin{eqnarray}
\partial_{l} V^{(1)} = \overline{P_1} (V^{(1)} * V^{(1)}) + O({V^{(1)}}^3)
\end{eqnarray}
To next order $O({V^{(1)}}^3)$ one needs to solve the coupled set:
\begin{eqnarray}
&& \partial_{l} V^{(1)} = \overline{P_1} ( V^{(1)} * V^{(1)} + V^{(1)} * V^{(2)} )
+  O({V^{(1)}}^4) \\
&& \partial_{l} V^{(2)} = (1-P_1) ( V^{(1)} * V^{(1)} )
\end{eqnarray}
The second equation is explicitly 
integrated which yields $V^{(2)}[V^{(1)}]$ which
is then substituted in the first equation, producing a closed equation
for $V^{(1)}$. This procedure can be extended to any order in $V^{(1)}$.
We now give the explicit calculation.

\subsection{RG equation up to order ${V^{(1)}}^3$}

\label{vcube}

To (lowest) order ${V^{(1)}}^2$, the beta function is local in $l$ as the 
modified Polchinski equation itself and reads
\begin{eqnarray}
\partial_{l} V^{(1)}(\phi,t)&=&\frac{1}{2} \int_{x_{1}x_{2}} 
\delta(t-\frac{x_{1}+x_{2}}{2}) 
e^{\partial^{1} G_l \partial^{2}}
\partial^{1} \partial G_l \partial^{2}
V^{(1)}(\phi_{1},x_{1}) |_{\phi_{1}=\phi}
V^{(1)}(\phi_{2},x_{2}) |_{\phi_{2}=\phi} 
\end{eqnarray}
up to terms of order ${\cal O}({V^{(1)}}^3)$.

To next order ${V^{(1)}}^3$, as explained above one first compute the 
bi-local operator as a function of $V^{(1)}$. Its flow
equation to the necessary order reads:
\begin{eqnarray}
\partial_{l}V^{(2)}_{l}(\phi_{1}\phi_{2},x_{1}x_{2}) &=& \frac{1}{2}
e^{\partial^{1} G_l \partial^{2}}
\partial^{1} \partial G_l \partial^{2}
V^{(1)}(\phi_{1},x_{1}) 
V^{(1)}(\phi_{2},x_{2}) \label{v2} \\
&&- \delta(x_{1}-x_{2}) \int_{y_{1}y_{2}} 
\delta(x_{1}-\frac{y_{1}+y_{2}}{2})e^{\partial^{1} G_l \partial^{2}}
\partial^{1} \partial G_l \partial^{2}
V^{(1)}(\phi_{1},y_{1}) 
V^{(1)}(\phi_{2},y_{2}) \nonumber
\end{eqnarray}
up to ${\cal O}({V^{(1)}}^3)$ terms. Integrating 
$\partial_{\mu}V^{(2)}_{\mu}(\phi_{1}\phi_{2},x_{1}x_{2})$ using 
(\ref{v2}) from $0$ to $l$ and substituting the result 
into the equation for $V^{(1)}$ one finds
the RG equation of the local part of the interaction to 
order ${V^{(1)}}^3$:
\begin{eqnarray}\label{polhat2}
&& \partial_{l} V^{(1)}(\phi,t)=\frac{1}{2} \int_{x_{1}x_{2}} 
\delta(t-\frac{x_{1}+x_{2}}{2})e^{\partial^{1} G_l \partial^{2}}
\partial^{1} \partial G_l \partial^{2}
V^{(1)}(\phi_{1},x_{1}) 
V^{(1)}(\phi_{2},x_{2}) \\
&&+\frac{1}{2}\int_{x_{1}x_{2}x_{3}} \delta(t-\frac{x_{1}+x_{2}+x_{3}}{3}) 
e^{\partial^{12}G_l\partial^{3}} \partial^{12} \partial G_l \partial^{3}
\left( \int_{0}^l d\mu \, \right. \nonumber \\
&&\left[ e^{\partial^{1}G_\mu\partial^{2}}
\partial^{1}\partial G_\mu\partial^{2} 
V^{(1)}_{\mu}(\phi_{1},x_{1}) |_{\phi_{1}=\phi}
V^{(1)}_{\mu}(\phi_{2},x_{2}) |_{\phi_{2}=\phi}\right. 
- \delta(x_{1}-x_{2}) \int_{y_{1}y_{2}} 
\delta(x_{1}-\frac{y_{1}+y_{2}}{2}) \nonumber \\
&&\left.\left.
e^{\partial^{1}G_\mu\partial^{2}}
\partial^{1}\partial G_\mu\partial^{2} 
V^{(1)}_{\mu}(\phi_{1},y_{1}) |_{\phi_{1}=\phi}
V^{(1)}_{\mu}(\phi_{2},y_{2}) |_{\phi_{2}=\phi}
\right] V^{(1)}_{l}  (\phi_{3},x_{3}) |_{\phi_{3}=\phi} \right)
\end{eqnarray}
up to ${\cal O}({V^{(1)}}^4)$ terms.

\subsection{Translation invariant theory and loop expansion}
\label{loopexpansion}

\begin{figure}[hbt]
\centerline{\epsfig{file=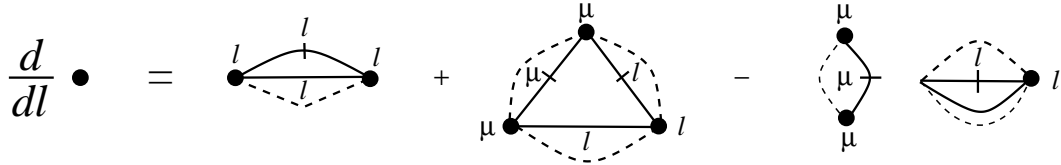,angle=0,width=14.cm}}
\caption{Graphical representation of the expansion of the modified
RG equation in powers of the local part up to ${\cal O}({V^{(1)}}^3)$.
The points represent the vertices $\hat{U}_l$, the broken solid lines
are propagators
on shell $\partial G$, the full solid lines are propagators
$G$, the indices $l$ or $\mu$ are indicated at 
each vertex and at each bond. While the solid lines are necessarily
present, additional
dashed lines appear in arbitrary number when performing
the loop expansion.}
\label{feynman}
\end{figure}

In a spatially translational invariant theory the local interaction
does not depend explicitly on the space variable $t$,
$V^{(1)}(\phi,t) = \hat{U}_l(\phi)$. 
The above formulas, when expanding the exponentials in a loop
expansion, possess a representation in terms of Feynman graphs as indicated 
in Fig.~\ref{feynman}. Interestingly, all one particle reducible graphs
vanish due to the property $\partial_l G_l^{q=0} = 0$ ($c'(0)=0$).
In addition, since each graph to order $\hat{U}^3$ possesses a 
counterpart with a minus sign which is the product of two 
(factorized) graphs with independent sets of loop integrations,
this automatically cancels all such (factorized) graphs.

The RG equation, at any given order in $\hat{U}_l$, can be further
expanded in the number of loops by expanding the exponentials 
in (\ref{polhat2}). Let us give the specific result for the case of
a diagonal gaussian part $G^q_{l,ij} = \delta_{ij} G^q_{l}$,
the generalization being straighforward.
To order ${\hat{U}_l}^3$ and up to two loops
we obtain from (\ref{polhat2})
the RG equation for $\hat{U}_l(\phi)$
as:
\begin{eqnarray}\label{polhat3}
\partial_{l} \hat{U}_l (\phi)&=&I^{D}_{l} D_{l}(\phi) + I^{F}_{l} F_{l}(\phi) +
\int_{0}^l d\mu \,\left( 
I^T_{l\mu} T_{l\mu}(\phi) + I^A_{l\mu} A_{l\mu}(\phi)
+I^{A'}_{l\mu} {A'}_{l\mu}(\phi) \right) 
\end{eqnarray}
up to $O({{\hat{U}_l}}^4)$ terms, 
where the contraction graphs are
\begin{eqnarray}
D_{l}(\phi) &=& \partial_{ij} \hat{U}_{l}(\phi) 
\partial_{ij}\hat{U}_{l}(\phi)\\
F_{l}(\phi) &=& \partial_{ijm} \hat{U}_{l}(\phi) 
\partial_{ijm}\hat{U}_{l}(\phi)\\
T_{l\mu }(\phi) &=& \partial_{ij} \hat{U}_{l}(\phi) 
\partial_{jm}\hat{U}_{\mu}(\phi) 
\partial_{mi}\hat{U}_{\mu}(\phi)\\
A_{l\mu }(\phi) &=& \partial_{ij} \hat{U}_{l}(\phi) 
\partial_{imn}\hat{U}_{\mu}(\phi)
\partial_{jmn}\hat{U}_{\mu}(\phi)\\
A'_{l\mu }(\phi) &=& \partial_{ij} \hat{U}_{\mu}(\phi)  
\partial_{imn}\hat{U}_{\mu}(\phi)
\partial_{jmn}\hat{U}_{l}(\phi)
\end{eqnarray}
and the momenta graphs read
\begin{eqnarray}
I^D_{l} &=& \frac{1}{2} \int_{q} \partial G_{l}^q G_{l}^q \\
I^F_{l} &=& \frac{1}{4} \int_{q_{1}q_{2}q_{3}} \delta_{q_{1}+q_{2}+q_{3}}
\partial G_{l}^{q_1} G_{l}^{q_2} G_{l}^{q_3} \\
I^T_{l\mu} &=& \int_{q} G_{l}^q  \partial G_{l}^q \partial G_{\mu}^q\\
I^A_{l\mu} &=& \int_{q_{1}q_{2}q_{3}}(\delta_{q_{1}+q_{2}+q_{3}}-
\delta_{q_{2}+q_{3}}) G_{l}^{q_{1}} \partial G_{l}^{q_{1}} 
G_{\mu}^{q_{2}} \partial G_{\mu}^{q_{3}}\\
I^{A'}_{l\mu} &=& \int_{q_{1}q_{2}q_{3}}\delta_{q_{1}+q_{2}+q_{3}}
(\frac{1}{2} \partial G_{l}^{q_{1}} G_{l}^{q_{2}} G_{l}^{q_{3}}  
\partial G_{\mu}^{q_{1}}+\partial G_{l}^{q_{2}} G_{l}^{q_{3}} 
G_{l}^{q_{1}} \partial G_{\mu}^{q_{1}})
\end{eqnarray}
Note that to two loops the RG flow is generically non local in $l$.
The values of the above integrals will be computed in \cite{us_part2}.

\subsection{RG equation for $U_l(\phi)$}
\label{rgu}

The systematic expansion in multilocal interactions can also be performed
directly on the functional ${\cal V}((\phi)$. The procedure parallels the previous section
and
its details are given in \cite{us_part2}. Here we
give only the result for a translationally invariant theory,
for the RG flow of $U_l(\phi)$ in the translationally invariant
case to order $U_l^3$. It reads:
\begin{eqnarray}
&& \partial U_l(\phi)=
-\frac{1}{2}(\partial G_{l})^{x=0}_{\alpha \beta }
\partial_{\alpha }\partial_{\beta } U_l(\phi) \\
&& -\frac{1}{4}\int_{x_{1}x_{2}}\delta (\frac{x_{1}+x_{2}}{2})
\partial^{12}(\partial G_{l }-\partial G^{x=0}_{l })\partial^{12} 
\int_{0}^{l}d\mu \,e^{-\frac{1}{2}\partial^{12}G_{l \mu }\partial^{12}}
\partial^{1}\partial G_\mu  \partial^{2}
U_\mu(\phi_{1})_{|\phi _{1}=\phi}
U_\mu(\phi_{2})_{|\phi _{2}=\phi} \nonumber \\
&& -\frac{1}{4}\int_{x_{1}x_{2}x_{3}}\delta (\frac{x_{1}+x_{2}+x_{3}}{3})
\int_{0}^{l}d\mu \nonumber \\
&& \left[\partial ^{123}(\partial G_{l }-\partial G^{x=0}_{l })\partial^{123}
e^{-\frac{1}{2}\partial ^{123}G_{l \mu }\partial ^{123}}
\partial^{12}\partial G_\mu  \partial^3 -\mbox{idem }x_{1}=x_{2}\equiv \frac{x_{1}+x_{2}}{2}\right] \nonumber \\
&& \int_{0}^\mu d\nu \, e^{-\frac{1}{2}\partial^{12}G_{\mu \nu}\partial^{12}}
\partial^{1}\partial G_{\nu }\partial^{2}
U_\nu(\phi_{1})_{|\phi _{1}=\phi}
U_\nu(\phi_{2})_{|\phi _{2}=\phi}
U_\mu(\phi_{3})_{|\phi _{3}=\phi} \nonumber
\end{eqnarray}
Note that the $\mbox{idem }x_{1}=x_{2}\equiv \frac{x_{1}+x_{2}}{2}$ applies only
in the square bracket to the arguments of the $G$'s. We have also defined:
\begin{eqnarray}
\partial^{1\dots p}\partial G \partial^{1\dots p}=\sum_{\alpha,\beta=1}^{p}
\partial
G^{x_{\alpha}x_{\beta}}_{ij }\partial_{i }^{\alpha}\partial_{j}^{\beta}
\end{eqnarray}
Again the 1PR diagrams are elimnated by construction since they have
one point $x_i$ on which one can integrate freely, producing 
a $\partial G^{q=0}$ which vanishes by construction.
The loop expansion of this formula will be detailed in 
\cite{us_part2}.

\subsection{Computation of the correlation functions via RG}

\label{correl}

The invariance of the generating functional $W (J)$ of the (connected) 
correlation functions with respect to $l$ is now used as a tool for 
computing the correlation functions of the initial model ${\cal
S}_{0}$. The expansion of $W (J)$ in powers of the running interaction
${\cal V}_{l} (\phi )$ reads formally to all orders:
\begin{eqnarray}
&& W(J)=\frac{1}{2}J:G:J + \sum_{m=1}^{+\infty} \kappa_m
\end{eqnarray}
where we have defined:
\begin{eqnarray}
&& \mu_n = \left[e^{J:G:G_{l}^{-1}:\phi}({\cal V}_{l} (\phi ))^n
\right]_{G_{l}} \\
&& Y = e^{-\frac{1}{2}J:G:G_{l}^{-1}:G:J} \\
&& \sum_{m=1}^{+\infty} \kappa_m x^m = \ln(1 + Y \sum_{n=1}^{+\infty}
\frac{(-x)^n}{n!} \mu_n)
\end{eqnarray}

Up to second order, this expansion reduces to:
\begin{eqnarray}\label{wjnohat}
&& W(J)=\frac{1}{2}J:G:J-e^{-\frac{1}{2}J:G:G_{l}^{-1}:G:J}
\left[e^{J:G:G_{l}^{-1}:\phi}{\cal V}_{l} (\phi ) \right]_{G_{l}} \\
&& + 
\frac{1}{2} e^{-\frac{1}{2}J:G:G_{l}^{-1}:G:J}
\left[ e^{J:G:G_{l}^{-1}:\phi}{\cal V}_{l} (\phi )^2 \right]_{G_{l}} 
- 
\frac{1}{2}
( e^{-\frac{1}{2} J:G:G_{l}^{-1}:G:J} 
\left[ e^{J:G:G_{l}^{-1}:\phi}{\cal V}_{l} (\phi ) \right]_{G_{l}} )^2
+ O({\cal V}_{l}^3) \nonumber
\end{eqnarray}

The interaction functional
$\hat{\cal V}_{l}(\phi)$ defined by (\ref{hatv}) naturally appears in the expansion of $W(J)$. 
Using the properties (\ref{gaussshift}) and $[{\cal A}(\phi)]_{G} = 
e^{\frac{1}{2}
\frac{\delta }{\delta \phi}:G:\frac{\delta }{\delta \phi}} {\cal
A}(0)$, one obtains:

\begin{eqnarray}
W(J)&=&\frac{1}{2}J:G:J - \hat{\cal V}_{l}(G:J) + \frac{1}{2}\left(
\widehat{{\cal V}^{2}_{l}}(G:J) - \hat{{\cal V}}^{2}_{l} (G:J) \right)
+ {\cal O}({\cal V}_{l}^3)
\end{eqnarray}
where $\widehat{{\cal V}^{2}_{l}}(\psi) = 
e^{\frac{\delta }{\delta \phi _{1}}:G_{l}:\frac{\delta }{\delta \phi
_{2}}} \hat{{\cal V}}_{l} (\phi _{1}) \hat{{\cal V}}_{l} (\phi _{2}) 
|_{\phi _{1}=\phi _{2}=\psi}$. 
On this expression, it becomes obvious that $W(J)$ is indeed $l$-independent
(order by order), as a consequence of the RG equation for $\hat{\cal V}_{l}(\phi)$.
As is clear from these formulae, all external legs of correlation functions
will carry the propagator $G$ while all internal legs will carry $G_l$. 

We must now distinguish between the two methods which consist in
performing the multi-local expansion on ${\cal V}_{l}(\phi)$,
$\hat{\cal V}_{l}(\phi)$ respectively. Before doing so, we give a
formula, in Fourier representation, which is valid in both cases:
\begin{eqnarray}
&& W(J)=\frac{1}{2} J:G:J-\int_{xK} \hat{U}^{K}_{l} e^{iK.(G:J)^{x}} \\
&& -\int_{xyKP} e^{iK.(G:J)^{x} + i P.(G:J)^{y}} \left(
\hat{V}^{K P}_{l}(x-y) - \frac{1}{2}
(e^{- K.G^{xy}_{l}.P} -1) \hat{U}^{K}_{l} \hat{U}^{P}_{l} \right)\\
&& \hat{U}^{K}_{l}= U^{K}_{l}e^{-\frac{1}{2} K.G^{x=0}_{l}.K} \\
&& \hat{V}^{KP}_{l}(x) = V^{K,P}_{l}(x)
e^{-\frac{1}{2}K.G^{x=0}_{l}.K - \frac{1}{2} P.G^{x=0}_{l}.P
- K.G^{x}_{l}.P}
\end{eqnarray}
the way to compute the functions $\hat{U}_l$ and $\hat{V}_l$ being
however different in each case. Inserting the corresponding formula
for $\hat{V}_l$ as a function of $\hat{U}_l$ yields expressions in
terms of $\hat{U}_l$ only, which we now give in each case (for variety, we also 
alternate between the -equivalent - field and Fourier representations). 

\subsubsection{Method with $\hat{\cal V}_{l}(\phi)$}

We start with the formalism using the multi-local expansion of 
$\hat {\cal V}_{l}$. One finds:

\begin{eqnarray}\label{wjofchecku}
&& W(J)=\frac{1}{2} J:G:J - \int_x \hat{U}_l((G:J)^x) 
+ \frac{1}{2} \int_{xy} \int_{l'>0} ~ 
[ \delta_{ll'} (e^{\partial_1 G_l^{x-y} \partial_2} -1 )
\\
&& - \theta_{ll'} \partial_{l'} ( e^{\partial^1 G_{l'}^{x-y} \partial^2} 
- \delta(x-y) \int_z e^{\partial_1 G_{l'}^{z} \partial_2} )
\hat{U}_{l'}(\phi_1)|_{\phi_1=(G:J)^x}  \hat{U}_{l'}(\phi_2)|_{\phi_2=(G:J)^y} ]
\end{eqnarray}
up to $O(U^3)$ terms. We denote $\delta_{ll'}=\delta(l-l')$ and $\theta_{ll'}=
\theta(l-l')$. From this formula one can compute all connected correlations
to $O(U^2)$. Let us give the self energy, defined as usual from the
two point function $C = G + \delta C$ as:
\begin{eqnarray}
\Sigma = C^{-1} - G^{-1} = - G^{-1} \delta C G^{-1} + 
G^{-1} \delta C G^{-1} \delta C G^{-1} + O(\delta C^3) 
\end{eqnarray}
It reads:
\begin{eqnarray}
&& \Sigma^{q=0}_{ij} = \partial_i \partial_j \hat{U}_l (0) 
- \int_x 
[ (\partial^1_i \partial^1_j + \partial_i^1 \partial_j^2)
(e^{\partial^1 G^{x}_{l} \partial^2} -1)
- \partial_i^1 \partial_j^2 
\partial^1 G^{x} \partial^2 ]
\hat{U}_{l}(\phi_1) \hat{U}_{l}(\phi_2)|_{\phi_i=0} \nonumber  \\
&& \Sigma^{q} - \Sigma^{q=0} = 
\int_x (e^{i q x} -1) \Sigma^x \\
&& \Sigma^x_{ij} = -
\partial_i^1 \partial_j^2 \int_{l'>0} 
[\delta_{ll'} (e^{\partial^1 G^{x}_{l} \partial^2}
-\partial^1 G^{x} \partial^2 -1)
- \theta_{ll'} \partial^1 \partial G^{x}_{l'} \partial^2
e^{\partial^1 G^{x}_{l'} \partial^2} ]
\hat{U}_{l'}(\phi_1) \hat{U}_{l'}(\phi_2)|_{\phi_i=0} \nonumber 
\end{eqnarray}
Note that it involves a term with a $G$ propagator.

\subsubsection{Method with ${\cal V}_{l}(\phi)$}

Inserting the multilocal expansion of ${\cal V}_{l}$, 
(\ref{wjnohat}) transforms into
an expansion in powers of the local interaction $U_l(\phi)$:
\begin{eqnarray}
&& W(J)=\frac{1}{2} J:G:J- \int_{x} \int_{K} \hat{U}^{K}_{l} e^{iK.G_x:J} 
+ \frac{1}{2} \int_{xy} \int_{KP} e^{iK.G_x:J + i P.G_y:J }[ 
\hat{U}^{K}_{l} \hat{U}^{P}_{l} 
(e^{- K.G^{x-y}_{l}.P} -1) \nonumber \\
&& -  \int_{0}^{l} dl' 
\hat{U}^{K}_{l'} \hat{U}^{P}_{l'} \partial_{l'} \left(
e^{- K.G^{x-y}_{l'}.P} - \delta(x-y) \int_z 
e^{- K.G^{z}_{l'}.P +
K.(G_l^z - G_l^0).P}  \right)]
+ O(U^3)
\end{eqnarray}
Using the RG equation for $U_{l}$, it is easily checked again that 
this expression is $l$-independent order by order.
From (\ref{wjofchecku}), one can compute the self energy $\Sigma^{q}$
of the theory. One gets to order $U_l^2$:
\begin{eqnarray}
&& \Sigma^{q=0}_{ij} = - \int_{K} K_i K_j U^{K}_{l} 
e^{-\frac{1}{2} K G_{l}^{x=0} K} + \int_{K P}
(K_i K_j A^{KP}_l(q=0) + K_i  P_j B^{KP}_l(q=0))\\
&& \Sigma^{q} - \Sigma^{q=0} = \int_{K P} K_i  P_j (B^{KP}_l(q) - B^{KP}_l(q=0)) 
\nonumber  
\end{eqnarray}
where
\begin{eqnarray}
&& A^{KP}_l(q) = \int_{l'>0} U^{K}_{l'} U^{P}_{l'} e^{-\frac{1}{2}
K G_{l'}^{x=0} K + P G_{l'}^{x=0} P } 
( \delta_{l-l'} \int_x e^{i q x} (e^{- K P G^{x}_{l}} -1) \\
&& B^{KP}_l(q) = \int_{l'>0} U^{K}_{l'} U^{P}_{l'} 
e^{-\frac{1}{2}
K G_{l'}^{x=0} K + P G_{l'}^{x=0} P } 
 ( \delta_{l-l'} \int_x e^{i q x} (e^{- K G^{x}_{l} P} +
K G^{x}_{0} P -1) \\
&&
+ \theta_{l-l'} \int_x (e^{i q x}- e^{K.(G^x_l - G^0_l).P}) K \partial G_{l'}^{x} P e^{- K G^{x}_{l'} P})
\nonumber 
\end{eqnarray}

\section{Dimensional reduction from graphs}
\label{fleurs}

\subsection{Perturbation theory}

In this appendix, we sketch diagrammatically how
the perturbation expansion in $R$ of the average
of any observable $A[u]$ at $T=0$ is the same as the one which would
be obtained in the Gaussian theory corresponding to a simple random
force.

Precisely, the actions
\begin{eqnarray}\label{actionbare}
{\cal S}[u]=\frac{1}{2T}\sum_a 
\int_{xy} u_a^x (\overline{G}^{-1})^{xy} u_a^y-\frac{1}{2T^{2}}\sum_{ab}\int_{x}R 
(u_{a}^{x}-u_{b}^{x})
\end{eqnarray}
and 
\begin{eqnarray}\label{actionrf}
{\cal S}_{\rm rf}[u]=\frac{1}{-2R''(0)}\sum_{ab} 
\int_{xy} u_a^x ((\overline{G}*\overline{G})^{-1})^{xy} u_b^y
\end{eqnarray}
where $(\overline{G}*\overline{G})^{x}=
\int_{z}\overline{G}^{x-z}\overline{G}^{z}$ yield the same results when
computing the average of any functional $A[u]$ of the replicated field
at $T=0$, e.g. $A[u]=\prod_{i} u_{a_{i}}^{x_{i}}$.

\begin{figure}[hbt]
\centerline{\epsfig{file=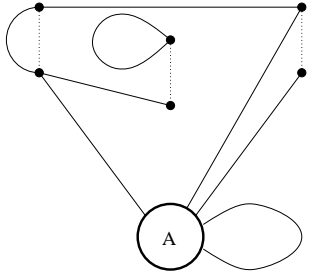,angle=0,width=4.cm}}
\caption{Typical graph contributing to the computation of the average of an 
observable $A$ to third order ($p=3$) in perturbation theory. Note here that 
$v=2p=6$, $t=1$, $k=3$, $e=4$, $c=3$, $l=1$.}
\label{example}
\end{figure}

To this aim, one first show that the perturbation expansion within
(\ref{actionbare}) is well-defined at $T=0$. We use a diagrammatics
with propagator
\begin{eqnarray}
\langle u^x_{a}u^y_{b}\rangle = T \delta_{ab} \overline{G}^{xy}
\end{eqnarray}
which conserves the replica index, and vertex
\begin{eqnarray}
-\frac{1}{2T^{2}} \sum_{ab}\int_{x} R (u^x_{a}-u^x_{b})
\end{eqnarray}
associated to {\it one} point in space but involving a summation over
{\it two} replica indices. Thus we choose to split the vertex into two
subvertices corresponding to each replica index.

For any graph occuring in the computation of $\langle A[u] R^p 
\rangle ^{c}_{\cal S}$, let us denote by $k$ the number of lines
connecting $A$ to the vertices (involving the extraction of $k$ legs
from $A$ :
$\partial_{u_{a_1}^{x_1}\ldots u_{a_k}^{x_k}}A[u]$). Let ${\cal K}$ be
the graph obtained by considering only the splitted vertices, the
propagators between them, forgetting the observable and the $k$ lines
attached to it. The graph ${\cal K}$ has $v=2 p$
subvertices. Contrarily to the initial graph with unsplitted vertices,
${\cal K}$ is not necessarily connected and is made of $c$ connected
components. To each one corresponds a replica index. 
If one of them is not connected to the
observable, i.e. if it does not inherit from a replica index contained
in $A$, then the summation over this index is free, giving a factor
$n$. Hence each connected component has to be linked to the observable
in order to survive the $n\to 0$ limit, which yields $k\geq c$.

Collecting the factors of $T$ in front of the initial graph 
(${\cal K}$, $A$, the $k$ propagators, and possibly $t$ tadpoles on
$A$), the power of $T$ is 
$t+e+k-v$ where $e$ is the number of propagators in ${\cal K}$.
Euler relation in ${\cal K}$ reads $v+l=c+e$ where $l$ is the number
of loops in ${\cal K}$. But since $k\geq c$ and $l\ge 0,  
t\ge 0$, one obtains that each graph is in factor of a non-negative
power of $T$. The existence of the $T=0$ perturbation theory 
is thus confirmed.

\begin{figure}[hbt]
\centerline{\epsfig{file=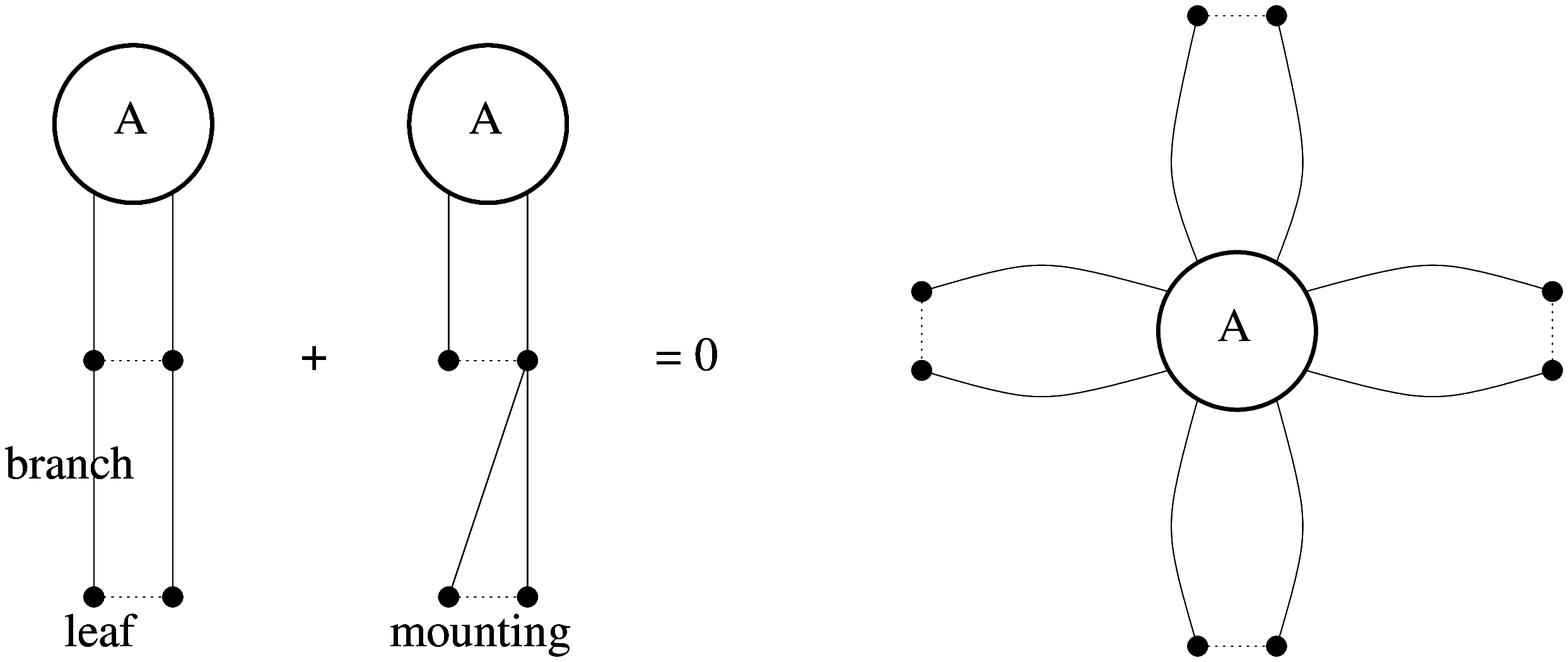,angle=0,width=8.cm}}
\caption{Vanishing contribution to the average of $A$ : the second graph is 
obtained by mounting a branch to the first graph. The only remaining
graphs are simple ``flowers''.}
\label{f:mount}
\end{figure}

The graphs which remain at $T=0$ have 
$t=0$, $l=0$, $k=c$, which means the following properties (i) 
their subgraph ${\cal K}$ has no loop, each of its connected
component is a tree, (ii) there is no tadpole on $A$, (iii) 
$A$ is linked to each connected component of ${\cal K}$  by one unique
propagator. This result is easily extended to a non-Gaussian disorder, which
possess higher cumulants of the general form (\ref{S(N)}).

The second part of the argument uses the property of translation
invariance in $u$ space of the disorder distribution,
on which the first part does not rely. Since each connected component
of ${\cal K}$ is linked to $A$, let us call {\it root} the point to
which it is attached. This provides a natural orientation to the {\it
branches} of the trees from root to {\it leaves}. If any point of ${\cal 
K}$ possess at least a branch going 
to the direction of the leaves, the
graph obtained by mounting this branch to the companion point (which
belongs to the same $R$ before splitting) has the opposite
value. One can convince oneself that such graphs can be grouped by mutually
cancelling
pairs. Thus the only graphs which survive to this mounting operation 
look like {\it flowers}, with $A$ at the center and
petals $R$ made of two propagators (see Fig. \ref{f:mount}).

The generalization including higher cumulants is straightforward, but
yields a non-Gaussian theory. The corresponding equivalent action for
computing observables is
\begin{eqnarray}\label{srf2}
{\cal S}_{\rm rf}[u]&=&\frac{1}{2T}\sum_a 
\int_{xy} u_a^x (\overline{G}^{-1})^{xy} u_a^y-\sum_{N\geq 2}
\frac{1}{N!T^{N}}\partial_{1\dots N}
S^{(N)} (0\dots 0)\int_{x} (\sum_{a}
u_{a}^{x})^{N}
\end{eqnarray}
Even if this is not obvious on (\ref{srf2}), this action possess
statistical tilt symmetry as can be checked thanks to $n\to 0$.

\subsection{Corrections to $R$}

The computation of the effective action $\Gamma $ (1PI) involves
corrections to the various cumulants of the disorder. At $T=0$, the 
graphs correcting a $N$ replica term ($N^{\rm th}$ cumulant) is made
of $N$ connected components, so that there exists a free sum over $N$
replica indices. The power of $T$ in front of such a cumulant has to
be $-N$. The graphs correcting $R$ with $p$ $R$'s are made of two connected
components ($c=2$), the power of $T$ is $e-v$ where $e$ is the number
of propagators and $v=2p$. Euler relation yields 
$e-v=l-c\geq -2$ with equality for $l=0$. Hence such graphs are made
of two trees.


Furthermore, the graphs such that the two points of a splitted $R$ are
connected to the same connected component and such that one of them is
connected to at least two branches vanish. This can be seen by
mounting one of these two branches on the companion point. 
Hence, if two points of a $R$ belong to the same connected
component, then each one is connected to a unique branch.
As a corrolary, the two points of an $R$ cannot be connected to each other
by a branch (since it would be impossible to connect this $R$ to the
rest of the diagram thanks to the argument above).

This considerably reduces the form of the possible corrections 
to $R$. These corrections obey in particular
\begin{equation}
\delta R = (\epsilon-4\zeta ) R + \zeta u R'+ 
\sum_{p>0} \left(\frac{d}{du} \right)^{4 (p-1)}R^{p}
\end{equation}
where the last term symbolically only means that
the $p^{\rm th}$ order term contains $4 (p-1)$ derivatives
(not that it is a total $4 (p-1)$ derivative). We
allowed for a field rescaling with exponent $\zeta $. To order $R^{3}$
the arguments above allow only for the following corrections
\begin{eqnarray}
\delta R = (\epsilon-4\zeta ) R + \zeta u R'
+ K (\frac{1}{2} R^{\prime \prime 2} - R'' R'' (0)) 
+ A (R''-R'' (0))R^{\prime \prime \prime 2} 
+ C (R''-R'' (0))^{2}R''''
\end{eqnarray}
with some constants $K$, $A$ and $C$, and valid only
for an analytic $R(u)$. 
In the periodic case ($\zeta =0$), the fixed point equation is easily
solved since there exists to any order in $\epsilon$ a fixed
point function of the form
\begin{equation}
R^{*} (u)=a+bu (1-u)+c\left(u (1-u) \right)^{2}
\end{equation}
where $a,b,c$ can be computed in series of $\epsilon $, once the
coefficients of the fixed point equation are known.
This is further examined in Sections \ref{totwoloops}.

\section{variational calculation}

\label{variational}

Here we sketch the derivation of the scaling function for
the confined interface using the replica variational method,
extending the explicit solution of Ref. \cite{mezard_parisi} to
a non zero mass. We use all notations of Ref. \cite{mezard_parisi}
and Ref. \cite{giamarchi_vortex_long}.

The disorder correlator for the random field problem studied
here corresponds to the case $\gamma= 1/2$ and $g=\sigma$
for the parameters of \cite{mezard_parisi}. Applying the
variational ansatz for $N=1$ components yields the 
function $\tilde{f}(x) = \hat{g} \sqrt{r_f^2 + x}$ which describes
the correlations, and $\hat{g} = \sigma \sqrt{2/\pi}$. We have
artifially extended the correlator to small scales, so as
to obtain a well defined $T=0$ limit. The large scale results however are
independent of the small scale details in the limit of small $m$.
The variational equations reads:
\begin{eqnarray}
(r_f^2 + 2 T B(u))^{3/2} = \hat{g} j_D ([\sigma](u) + m^2)^{-\epsilon/2} 
\end{eqnarray}
with $j_D = \int_k (k^2 + 1)^{-2}= \Gamma(2-d/2)/(4 \pi)^(d/2)$ 
and the equation for the breakpoint 
(see Ref. \cite{giamarchi_vortex_long}) is:
$(r_f^2 + 2 T B(u_c))^{3/2} = \hat{g} j_D (\Sigma_c + m^2)^{-\epsilon/2}$.
This yields the solution:
\begin{eqnarray}
&& [\sigma](u) + m^2 = A u^{2/\theta} = \Sigma_c (u/u_c)^{2/\theta}
\quad u^*<u<u_c \\
&& \sigma(u)=\sigma(0)=A^{\theta/2} \frac{2}{2 - \theta} m^{2-\theta} 
\quad u < u^*
\end{eqnarray}
and $[\sigma](u)=\Sigma_c$ for $u>u_c$. Here $\theta = \frac{6-\epsilon}{3}$
is the free energy fluctuation exponent and
$A = (\frac{\epsilon}{6 T} \hat{g}^{2/3} j_D^{-1/3} )^{2/\theta}$.
This solution allows to compute the small $q$ behaviour of the
correlation (for small $m \sim q$) asL

\begin{eqnarray}
&& \overline{u_q u_{-q} } = \frac{T}{q^2 + m^2} ( 1 +
\int_{u_c (m^2/\Sigma_c)^{\theta/2}}^{u_c} \frac{du}{u^2} 
\frac{\Sigma_c (u/u_c)^{2/\theta} - m^2}{q^2 + \Sigma_c (u/u_c)^{2/\theta}} 
+ \sigma(0) \frac{1}{q^2 + m^2} ) 
\end{eqnarray}
which yields the large scale result \ref{rfimmp} given in the text.
In $D=0$ one recovers the result of \cite{mp2}
\begin{eqnarray}
\overline{u^2} = \frac{3}{(4 \pi)^{1/3}} m^{-8/3} \sigma^{2/3}
\end{eqnarray}

\section{FRG to two loops}\label{totwoloops}

\subsection{Method with ${\cal V}$ }

The exact RG equation to order $U^3$ given in \ref{rgu} when expanded to
two loops yields the following finite temperature RG equation for
$\tilde{R}_l(u)$ at large $l$ (the derivation is detailed in \cite{us_part2}):

\begin{eqnarray}
&& \partial_{l}\tilde{R}_{l}=\epsilon \tilde{R}_l
+ \hat{T}_l \tilde{R}_{l}'' 
+ K_{l \mu} \left(\frac{1}{2}\tilde{R}_{\mu }''
(u)^{2}-\tilde{R}_{\mu }'' (0)\tilde{R}_{\mu }'' (u) 
\right)  \\ 
&& + K^A (\tilde{R}'' -\tilde{R}''(0)) \tilde{R}'''^{2}
+ K^C (\tilde{R}'' - \tilde{R}''(0))^{2}\tilde{R}'''' \\
&& + K^S \partial_{12}\tilde{S}_l (u,u,0)
+ \hat{T}_l K^E_{l \mu} \left(\tilde{R}_{\mu }''''  (\tilde{R}_{\mu }''
(u)-\tilde{R}_{\mu }'' 
(0))-\tilde{R}''''_{\mu } (0)\tilde{R}_{\mu }''  \right) 
+ \hat{T}_l K^F_{l \mu} \tilde{R}_{\mu}^{\prime \prime \prime 2}
\label{frg2loopt}
\end{eqnarray}
In this formula all terms of order $R^2$ and higher are retarded, and
integrals $\int_0^l d\mu$ are understood. For the $R^3$ terms we have ommitted
the retardation integrals (which involve an additional integral $\int_0^\mu d\nu$)
because near the fixed point they can be replaced by a single number. 
The feedback of the three replica term is through its partial derivatives.
This three replica term satisfies its own RG equation given in \cite{us_part2}.
The precise values of all coefficients are detailed in \cite{us_part2}.

To obtain the large $l$ limit of this equation and thus the fixed
point equation 
we use the fact that $\hat{T}_l \to 0$ and the property
(\ref{important}) which we need 
only to lowest order. Integrating out the third cumulant RG equation
(which reaches 
the fixed point value (\ref{3rep})) 
and inserting into (\ref{frg2loopt}) yields the 
FRG fixed point equation (for simplicity in the periodic case):
\begin{eqnarray}
&& 0 = \epsilon  \tilde{R}^* + 
\frac{1}{2} \tilde{R}^{* \prime \prime 2}  - 
\tilde{R}^{* \prime \prime}(0) \tilde{R}^{* \prime \prime} \\
&& 
+ (\frac{1}{2} + \overline{K}^C) (\tilde{R}^{* \prime \prime} - \tilde{R}^{* \prime \prime}(0))
\tilde{R}^{* \prime \prime \prime 2} - \frac{1}{2} \tilde{R}^{* \prime
\prime \prime}(0^+)^2 
 \tilde{R}^{* \prime \prime}
+ \overline{K}^C (\tilde{R}^{* \prime \prime} - \tilde{R}^{* \prime
\prime}(0))^2  
\tilde{R}^{* \prime \prime \prime \prime}
\end{eqnarray}
where only one coefficient, $\overline{K}^C$, is non universal. We
have absorbed the exact one loop coefficient ${\sf K}$ obtained in (\ref{univers}) 
in $R$ (whose first ${\cal O}(\epsilon)$ correction is non universal). 
In terms of the cutoff function $c(s)=\int_a \hat{c}(a)
e^{-a s}$ (or its expression as a sum of exponentials) one has:
\begin{eqnarray}
&&  {\sf K} = \frac{1}{2 (2 \pi)^{d/2}} 
[ 1 + \epsilon ( \frac{1}{2} - \int_{a} \hat{c}(a) \ln a
+ \frac{3}{2} \int_{ab} \hat{c}(a) \hat{c}(b) \ln(a+b) )] \nonumber \\
&& \overline{K}^C = - \int_{a} \hat{c}(a) \ln a + \int_{ab} \hat{c}(a)
\hat{c}(b)
\ln(a+b) = \int_{0}^{\infty }ds\,\frac{c (s) (1-c (s))}{s}
\end{eqnarray}
to the desired order in $\epsilon$.
This equation has a fixed point solution:
\begin{eqnarray}
&& \tilde{R}^{*}(0) = {\sf K}^{-1} [
\epsilon (\frac{1}{2592} - \frac{1}{72} u^2 (1-u)^2 )
+ \\
&&
\epsilon^2 (\frac{1 + 10 \overline{K}^C}{7776} - \frac{\overline{K}^C}{216}
u(1-u) - \frac{1 + 3 \overline{K}^C}{108} u^2 (1-u)^2) ]
\end{eqnarray}
Note that this solution, for $\overline{K}^C=0$ (i.e hard cutoff) does not 
contain a term with $R'(0^+) \neq 0$ (``supercusp'')

\subsection{method with $\hat{\cal V}$ }

The method using the Wick ordered functional can also be used. Using the
RG equation to third order in $\hat{U}$ and two loops (\ref{polhat3})
one finds the equation for the corresponding two replica part of
$\hat{U}$:

\begin{eqnarray}
&& \partial_{l}\tilde{R}_{l}=(\epsilon - 4 \zeta) \tilde{R}_l
+ \zeta u \tilde{R}_l' 
+ K \left(\frac{1}{2}\tilde{R}_{l}''
(u)^{2}-\tilde{R}_{l }'' (0)\tilde{R}_{l}'' (u) 
\right)  \\ 
&& + K^A (\tilde{R}'' -\tilde{R}''(0)) \tilde{R}'''^{2} 
+ \hat{T}_l K^F_{l} \tilde{R}_{l}^{\prime \prime \prime 2}
\label{frg2loophat}
\end{eqnarray}
with:
\begin{eqnarray}
&& K = -\frac{4}{T^2} I^D_l \Lambda_l^{\epsilon} \\
&& K^A_{\l \mu} = \frac{4}{T^4} (I^A_{\l \mu}  + I^{A'}_{\l \mu}) \\
&& K^F_{l \mu} = -\frac{6}{T^2} I^F_l \Lambda_l^{\epsilon}
\end{eqnarray}
where the integrals have been defined in Section \ref{loopexpansion}.
Calculation shows that the constant $K^A = 1/2$
(to lowest order in $\epsilon$) independent of the cutoff function $c(s)$.
The coefficient of this term is in agreement with \cite{larkin_2loops,kay_dynamics}.
The analysis of the boundary layer is more intricate in this formulation
\cite{us_part2}.


\begin{thebibliography}{9}
\bibitem{analyticity}
The condition (\ref{examplecutoff},\ref{ejphi}), 
if strictly imposed, implies that $c (x)$ is non-analytic. 
However, a non-analytic cutoff can be obtained by a suitable limit of
analytic functions.
\bibitem{footnote3}
One has $a_{2 n} = \sum_{p \ge n} \frac{1}{(p-n)!} g_{2p} 
A_{np} (G^{x=0}_l/2)^{p-n}$
with $A_{np}=\prod_{k=n}^{p-1} (n+2 k)/(1+2 k)$ and 
$\tilde{a}_{2 n} 
= \sum_{p \ge n} \frac{1}{(p-n)!} \tilde{g}_{2p} A_{np} (K/2)^{p-n}$
with $K=\Lambda_l^{2-D} G^{x=0}_l = S_D \int_{s>0} (2 s)^{-\epsilon/2} c(s)$.
\end{thebibliography}
\end{document}